\documentclass[12pt,preprint]{aastex}
\shorttitle{NEAR INFRARED BACKGROUND \& EARLY STARS}
\shortauthors{FERNANDEZ \& KOMATSU}
\begin{document}
\title{%
  THE COSMIC NEAR INFRARED BACKGROUND: REMNANT LIGHT FROM EARLY STARS
}%
\author{%
  Elizabeth R. Fernandez and Eiichiro Komatsu
}%
\affil{%
  Department of Astronomy, University of Texas at Austin,
  1 University Station, C1400, Austin, TX 78712
}%
\email{%
  beth@astro.as.utexas.edu
}%
\begin{abstract}
  The redshifted ultraviolet light from early stars at $z\sim 10$ 
  contributes to the cosmic near infrared background. We present detailed
  calculations of its spectrum with various assumptions about metallicity 
  and mass spectrum of early stars. 
  We show that if the near infrared background has a stellar origin, 
  metal-free stars are {\it not} the only explanation of the excess
  near infrared background; stars with metals
  (e.g. $Z=1/50Z_{\sun}$)  can produce the same amount of background intensity as the
  metal-free  stars.
  We quantitatively show that the predicted average 
  intensity at 1--2~${\mu\rm m}$ is essentially determined by the
  efficiency of nuclear burning in stars, 
  which is not very sensitive to metallicity.
  We predict $\nu I_\nu/\dot{\rho}_*\simeq 4-8~{\rm nW~m^{-2}~sr^{-1}}$,
  where $\dot{\rho}_*$ is
  the mean star formation rate at $z=7-15$ (in units 
  of ${\rm M_\sun~yr^{-1}~Mpc^{-3}}$) for stars more massive than $5~M_\sun$.
  On the other hand, since we have very little
  knowledge about the form of mass spectrum of early stars, uncertainty in the average 
  intensity due to the mass spectrum could be large.
  An accurate determination of the near infrared background allows us to probe 
  formation history of early stars, which is difficult to constrain by other means.
  While the star formation rate at $z=7-15$ inferred from the current data is 
  significantly higher than the local rate at $z<5$, it does not rule out 
  the stellar origin of the cosmic near infrared background. In addition, we show that a reasonable initial mass function, coupled with this star formation rate, does not over-produce metals in the universe in most cases, and may produce as little as less than 1 \% of the metals observed in the universe today.
\end{abstract}
\keywords{cosmology: theory --- diffuse radiation --- infrared: galaxies}
\section{INTRODUCTION}
\label{sec:introduction}

When and how was the universe reionized? 
These questions have actively been studied almost purely by theoretical
means (see \citet{barkana/loeb:2001,bromm/larson:2004,ciardi/ferrara:2005} 
for recent reviews),
as currently there are only a very few observational probes of the 
epoch of reionization: the Gunn--Peterson test 
\citep{gunn/peterson:1965,becker/etal:2001},
polarized light of the cosmic microwave background on large angular 
scales \citep{zaldarriaga:1997,kaplinghat/etal:2003},
which has been detected by the {\sl Wilkinson Microwave Anisotropy Probe}
({\sl WMAP}) \citep{kogut/etal:2003},
and temperature of the intergalactic medium \citep{hui/haiman:2003}.
Over the next decades, we are hoping to
detect the first sources of light directly with the next
generation of space telescopes, such as the {\sl 
James--Webb Space Telescope} ({\sl JWST}).
More ambitious are mapping observations and measurements of the 
power spectrum of fluctuations of the 21-cm line background from neutral 
hydrogen atoms during reionization
\citep{ciardi/madau:2003,furlanetto/sokasian/hernquist:2004}
or even prior to reionization 
\citep{scott/rees:1990,madau/meiksin/rees:1997,tozzi/etal:2000,
iliev/etal:2002}, which offer very powerful probes of detailed history
of the cosmic reionization.

It has been pointed out that the mean intensity 
\citep{santos/bromm/kamionkowski:2002,salvaterra/ferrara:2003,cooray/yoshida:2004,madau/silk:2005}
as well as fluctuations 
\citep{magliocchetti/salvaterra/ferrara:2003,kashlinsky/etal:2004,cooray/etal:2004}
of the near infrared background potentially offer yet another
window to the epoch of reionization.
The logic is very simple: suppose that most of reionization occurred
at, say, $z=9$. The ultraviolet photons ($\lambda\sim 1000~$\AA)
produced at $z=9$ during reionization will then be redshifted to the 
near infrared regime ($\lambda\sim 1~\mu{\rm m}$). 
In other words, a fraction of the near infrared background 
(whether or not observable) {\it must come from the epoch of reionization},
and there is no question about the existence of the signal.
(Of course the existence of the signal does not immediately imply 
that the signal is actually significant.)
It is therefore extremely important to understand the near infrared
background in the context of redshifted UV photons 
and examine to what extent it is relevant to and useful for understanding
the physics of cosmic reionization.

Has it been detected? All of the theoretical proposals were essentially 
motivated by the current measurements of the near infrared background, 
which suggests the existence of an isotropic background after subtraction of 
the zodiacal emission
\citep{dwek/arendt:1998,gorjian/wright/chary:2000,wright/reese:2000,wright:2001,cambresy/etal:2001,matsumoto/etal:2005, kashlinsky:2005}.
Since the zodiacal emission is $\sim 3$ times as large as the inferred 
isotropic component, one should generally be careful when interpreting
the data. Although the inferred background is isotropic at the first order, 
significant fluctuations still remain at the level of $\Delta I/I\sim 1/4$
\citep{kashlinsky/odenwald:2000,kashlinsky/etal:2002,matsumoto/etal:2005},
which requires further explanations.
The most intriguing feature of the current observational data is that
the inferred background seems too large to be accounted for by
the integrated light from galaxies (\citet{totani/etal:2001}, see also Figure~12 of \citet{matsumoto/etal:2005}
and references therein for the compilation of the galactic contribution).
It is thus tempting to speculate that the bulk of the near infrared background
(aside from the zodiacal light) actually comes from stellar sources at 
the epoch of reionization.

In this paper, we carefully examine the near infrared background from early stars.
While our approach is similar to that of \citet{santos/bromm/kamionkowski:2002},
which has been adopted by most of the subsequent work,
our goal is to 
(1) simplify physics and improve calculations,
(2) explore different metallicity and initial mass spectra,
(3) understand robustness of theoretical predictions, and
(4) provide a simple relation between the cosmic near infrared background 
and star formation rate.
The focus of this paper is the mean intensity: we will discuss fluctuations 
in the forthcoming paper (Fernandez et al., in preparation).
This paper is organized as follows. 
In \S~\ref{sec:formalism}, we develop the basic formalism and summarize
relevant emission processes such as stellar emission and reprocessed light, 
the latter including Lyman-$\alpha$, two-photon, free-free and free-bound emission.
In \S~\ref{sec:energy}, we examine energy spectra of various emission 
processes from early stars. In particular we explore differences in
the energy spectrum between various assumptions about metallicity and 
initial mass spectrum of early stars.
In \S~\ref{sec:irb}, we calculate the spectrum of the cosmic near 
infrared background. In \S~\ref{sec:sfr},
we compare the prediction to the current observational data and discuss
implications for the star formation rate at $z=7-15$. 
In \S~\ref{sec:metals}, constraints from metal production from the first stars are calculated.
In \S~\ref{sec:remarks} we discuss the other constraints from
a collapse fraction of dark matter halos.
We conclude in \S~\ref{sec:conclusion}.

\section{STELLAR EMISSION AND REPROCESSED LIGHT}
\label{sec:formalism}

\subsection{Basic Formalism}
\label{sec:basics}

We calculate the background intensity, $I_{\nu}$, as \citep{peacock:1999}
\begin{equation}
 I_{\nu} =
 \frac{c}{4\pi} 
 \int 
 \frac{dz\, p([1+z]\nu, z)}{H(z) (1+z)},
\label{eq-inu-generic}
\end{equation}
where $\nu$ is an observed frequency (which is in the near
infrared band: for $\lambda=3-1~\mu{\rm m}$, say, 
$\nu=100-300~{\rm THz}$ or $h\nu=0.414-1.24~{\rm eV}$), 
$H(z)$ is the expansion rate at
redshift $z$ ($dt/dz = -[H(z)(1+z)]^{-1}$), and 
$p(\nu, z)$ is the volume emissivity 
in units of energy per unit time, unit frequency and unit
{\it comoving} volume. There are several contributions to
the emissivity. One is the continuum emission from stars
themselves, $p^*$, which is nearly a black body spectrum, 
and the others are reprocessed light of ionizing radiation: 
a star ionizes neutral gas in its neighborhood and a series of 
recombination lines, $p^{line}$, emerge.
The ionized gas (or nebula) also emits free-free and free-bound
continuum emission, $p^{cont}$, as well as two-photon emission, $p^{2\gamma}$.
In Appendix~\ref{sec:emissivity}, we derive the formula for the volume
emissivity as (Eq.~[\ref{eq:p}])
\begin{equation}
  p(\nu,z)  
= \dot{\rho}_*(z)c^2 \sum_\alpha \langle \epsilon^\alpha_{\nu}\rangle,
\end{equation}
where 
\begin{equation}
  \langle \epsilon^\alpha_{\nu} \rangle
  \equiv
  \frac1{m_*}\int dm~mf(m)
  \left[\frac{\overline{L}^\alpha_\nu(m)\tau(m)}{mc^2}\right],
  \label{eq:avg}
\end{equation}
and $f(m)$ is a mass spectrum (specified later 
in \S~\ref{sec:imf}; for the precise definition, see Appendix~\ref{sec:emissivity}),
and $m_*$ is the mean stellar mass (Eq.~[\ref{eq:mstar}]).
It is important to note that this formula has been derived assuming
that the stellar main-sequence lifetime, $\tau(m)$, is shorter
than the Hubble time, and corrected for dead stars which do not contribute
to the volume emissivity.
Here, $\overline{L}_\nu^\alpha(m)$ is a time-averaged luminosity (over the main-sequence lifetime) in frequency
interval [$\nu$, $\nu+d\nu$] for a radiative process of $\alpha$, and 
$\langle\epsilon^\alpha_{\nu}\rangle$ is the key dimensionless quantity which 
represents a ratio of the {\it mass-weighted} average\footnote{Throughout this paper, 
we shall use $\langle\rangle$ to denote the mass-weighted average.} 
of total radiative energy (including stellar emission and
reprocessed light) to the stellar rest mass energy, in unit frequency 
interval.
In other words, $\langle\epsilon^\alpha_{\nu}\rangle$ represents the mass-weighted
mean radiative efficiency of stars.
This formulation is useful as one can immediately see
that each contribution is simply given by
the star formation rate (which depends on $z$) 
and a typical radiative efficiency (which does not
depend on $z$).
While $\dot{\rho}_*$ is very uncertain and will be constrained by a 
comparison to the observational data, one can calculate
$\langle \epsilon^\alpha_{\nu}\rangle$ robustly for a 
given population of stars using simple physical arguments.
It will be shown in the subsequent sections that 
$\nu\langle \epsilon^\alpha_{\nu}\rangle$ is always of order  
$10^{-3}$ when averaged over the main-sequence lifetime, which can be understood with simple energetics.
Initially, energy must be generated by nuclear burning in stars.
While the rest mass energy of $1~M_\sun$ is as big as 
$1.8\times 10^{54}~{\rm erg}$, only a fraction will go into radiative energy.
For example, in the Sun only 0.07\% of the rest mass energy is converted
to radiative energy over its main-sequence lifetime.
The nuclear burning efficiency\footnote{By ``nuclear burning efficiency'',
we mean the bolometric energy of stellar emission before absorption per stellar rest mass 
energy, $\overline{L}^*_{bol}\tau/mc^2$.} depends on stellar mass only weakly
at large masses. 
Our detailed calculations below confirm
this simple argument, and thus the uncertainty in the 
predicted amplitude of radiative efficiency is small for a given 
mass spectrum of stars.

Using the expected radiative efficiency of stars, we 
obtain 
\begin{equation}
  \nu I_\nu = 13.1~{\rm nW~m^{-2}~sr^{-1}}
 \int 
 \frac{d y~10^2}{y^2{\cal E}(y)}
\left[
\frac{\dot{\rho}_*(y)}{M_\sun~{\rm yr^{-1}~Mpc^{-3}}}
\right]
\sum_\alpha\frac{(\nu y) 
\langle \epsilon^\alpha_{(\nu y)} \rangle}{10^{-3}},
\label{eq-inu-order}
\end{equation}
where $y\equiv 1+z$ and
\begin{equation}
 {\cal E}(y)\equiv
  h\sqrt{\Omega_m y^3+\Omega_\Lambda}
  \simeq 11.83\left(\frac{\Omega_mh^2}{0.14}\right)^{1/2}
\left(\frac{y}{10}\right)^{3/2},
\end{equation}
for the redshift range of interest.
Thus, without any detailed calculations, one can show that 
the cosmic near infrared background from early stars at $z\sim 10$
should be approximately given by
$\nu I_\nu/\dot{\rho}_*\sim 10~{\rm nW~m^{-2}~sr^{-1}}$ where
$\dot{\rho}_*$ is in units of $M_\sun~{\rm yr^{-1}~Mpc^{-3}}$.

\subsection{Stellar Contribution}
\label{sec:stellar}

To simplify calculations, we assume that 
the stellar spectrum is a black body with
the Lyman continuum photons completely absorbed:
\begin{equation}
  \label{eq:stellar}
  \overline{L}^{*}_\nu(m) = \left\{
    \begin{array}{ll} 
  4\pi R_*^2(m) B_\nu[T_{eff}(m)], & h\nu<13.6~{\rm eV} \\
  0, &   h\nu\ge 13.6~{\rm eV}
  \end{array}       
  \right.
\end{equation}
where $R_*$ is a stellar radius and $T_{eff}$ is the effective 
temperature, and 
$B_\nu[T_{eff}(m)]$ is a black body spectrum given by
\begin{equation}
B_\nu[T_{eff}(m)]= \frac{2h\nu^3/c^2}{\exp(h\nu/kT)-1}.
\end{equation}
Note that the stellar spectrum (before absorption) above
13.6~eV (which determines the number of hydrogen-ionizing
photons) is significantly different from a black body;
thus, we do not use a black-body spectrum to calculate the 
number of ionizing photons, but use more detailed calculations
by \citet{schaerer:2002} (see \S~\ref{sec:metallicity}).
The stellar spectrum just below 13.6~eV is also different
from a black body because of a cluster of absorption lines of 
Lyman series; however, we ignore this effect and keep our calculations 
as simple as possible. (One can always use a more precise stellar 
spectrum for a better accuracy.)

\subsection{Free-free and Free-bound Contribution}
\label{sec:nebular}

The free-free and free-bound continuum luminosity is given by
\begin{equation}
\label{eq:ffbf_org}
  \overline{L}^{cont}_\nu(m) = \frac{\varepsilon_\nu\overline{Q}_H(m)}{n_e n_p\alpha_B},
\end{equation}
where $\overline{Q}_H$ is a time-averaged production rate of hydrogen ionizing photons
(the average number of ionizing photons produced per unit time),
$n_e$ and $n_p$ are the number density of electrons and protons, respectively, 
$\alpha_B$ is the case-B recombination coefficient for hydrogen
(Eq.~[5-14] of \citet{spitzer:1978} with $n=2$ and $Z=1$) given by
\begin{equation}
  \alpha_B =  \frac{2.06\times 10^{-11}}{T_g^{1/2}}\phi_2(T_g)~{\rm cm^3~s^{-1}},
  \label{eq:caseB}
\end{equation}
and $\phi_2(T_g)$ is a dimensionless function of temperature
tabulated in Table~5.2 of \citet{spitzer:1978}.
Here, $T_g$ denotes gas temperature in units of Kelvin.
In principle, to calculate $T_g$ one has to equate the energy gain 
and loss to find out equilibrium temperature.
While $T_g$ varies depending on stellar temperature (or hardness of a stellar
spectrum which determines photo-heating), we shall assume $T_g=20000~{\rm K}$ 
regardless of the stellar temperature, which should be a good approximation
for our purposes. For this temperature we find  
$\phi_2(T_g=20000~{\rm K})\simeq 1$.

The quantity $\overline{Q}_H(m)/(n_e n_p\alpha_B)$ is the volume of the 
Str\"omgren sphere (see text below Eq.~[\ref{eq:ffbf}]), and
$\varepsilon_\nu$ is the total volume emissivity including 
free-free and free-bound emission (Eq.~[6.22] of 
\citet{dopita/sutherland:2002}):
\begin{equation}
  \varepsilon_\nu = 4\pi n_e n_p\gamma_c
\frac{e^{-h\nu/kT_g}}{T_g^{1/2}}~{\rm erg~cm^{-3}~s^{-1}~Hz^{-1}},
\end{equation}
where $\gamma_c$ is the continuum emission coefficient
including free-free and free-bound emission:
\begin{equation}
  \gamma_c \equiv f_k\left[\overline{g}_{ff}
  + \sum^{\infty}_{n=2}\frac{x_ne^{x_n}}{n}g_{fb}(n)\right],
\label{eq:gammac}
\end{equation}
where $x_n\equiv {Ry}/(kT_gn^2)$, $\overline{g}_{ff}$ and $g_{bf}(n)$ 
are the Gaunt factors for free-free 
(which is thermally averaged) and free-bound emission, respectively,
and $f_k$ is the collection of physical constants
which has a numerical value of $5.44\times 10^{-39}$ in cgs units.
Note that we have ignored the helium contribution and assumed 
complete ionization for computing $\gamma_c$.
As a free-bound transition to $n=1$ will not be considered in
the case-B recombination, the summation is taken from $n=2$. 
(This is because all photons that recombine directly to $n=1$ 
are strongly absorbed by neighboring hydrogen atoms and immediately 
ionize them.) We then obtain
\begin{equation}
  \overline{L}^{cont}_\nu(m) 
  \simeq
  3.32\times 10^{22}~{\rm erg~s^{-1}~Hz^{-1}}
  \left[\frac{\overline{Q}_H(m)}{10^{49}~{\rm s}^{-1}}\right]
 \left[\overline{g}_{ff}+\frac{Ry}{kT_g}\sum_{n=2}^\infty
 \frac{e^{Ry/(kT_gn^2)}}{n^3}g_{fb}(n)\right]
 \phi_2(T_g) e^{-h\nu/kT_g}.
\label{eq:ffbf}
\end{equation}
The continuum luminosity does not depend on the number density of electrons
or protons. This is an immediate consequence of the Str\"omgren sphere: 
while the higher number density implies the larger emissivity, it also implies
the larger recombination rate and the smaller ionized region. These two effects
cancel out exactly, making luminosity independent of the number density.
Of course, this approximation breaks down in the intergalactic medium 
(outside of halos) in which ionization fronts do not fill the Str\"omgren 
sphere \citep{shapiro/giroux:1987}.
Our calculation assuming the Str\"omgren sphere is accurate
if the bulk of luminosity comes from nebulae around stars inside the host
halos, while it should give a robust upper limit on free-free and free-bound luminosity
otherwise.

Finally, we need to compute the Gaunt factors. For the parameter space we are 
interested in, 
\begin{eqnarray}
u&\equiv& \frac{h\nu(1+z)}{kT_g}={\cal O}(10)\frac{1+z}{10},\\
\gamma^2&\equiv& \frac{Ry}{kT_g}={\cal O}(10),
\end{eqnarray}
both Gaunt factors are approximately constant and given by 
\citep{karzas/latter:1961}
\begin{eqnarray}
\overline{g}_{ff} &\simeq& 1.1,\\
g_{fb}(n) &\simeq& 1.05,
\end{eqnarray}
which are accurate to within 10\%.

\subsection{Line Contribution}
\label{sec:line}

The line luminosity is given by
\begin{equation}
  \overline{L}^{line}_\nu(m) = \sum_i h\nu_i \phi_i(\nu-\nu_i) \dot{n}_i(m),
\end{equation}
where $\phi_i(\nu-\nu_i)$ is the line profile 
and $\dot{n}_i$ is a photon production rate at a line $i$.
Since the intergalactic medium is optically thick to
the Lyman continuum photons before the end of reionization,
every single hydrogen-ionizing photon will be absorbed and
converted to line emission; thus, the line contribution should be 
proportional to a production rate of hydrogen-ionizing photons, 
$Q_H$, as
\begin{equation}
  \dot{n}_i(m) = f_i \overline{Q}_H(m),
\end{equation}
where $f_i$ is a fraction of ionizing photons which are converted
to a line $i$. 

Which lines are important to the near infrared background?
The Lyman series photons are in right bands; however,
they are strongly absorbed and eventually converted to 
other lines. One exception is the Lyman-$\alpha$ photons:
while they are also strongly absorbed, they are re-emitted back again
in Lyman-$\alpha$. Therefore the net effect is that 
the Lyman-$\alpha$ photons are not destroyed (in the absence of dust)
but merely
scattered. \citet{loeb/rybicki:1999} have shown that 
as the universe expands the Lyman-$\alpha$ photons are eventually 
``redshifted out'' of scattering and escape freely.
The Balmer series photons (and others) have too low an energy to be
relevant to the near infrared background (a direct recombination
to $n=2$ results in a line at 3.4~eV or 3648~\AA, which will be 
redshifted to $\lambda\gtrsim 3~\mu{\rm m}$ and is thus irrelevant).
Therefore, we consider only Lyman-$\alpha$ photons:
\begin{eqnarray}
\nonumber
  \overline{L}^{line}_\nu(m) 
&=& 
f_{ly\alpha}h\nu_{ly\alpha} \phi(\nu-\nu_{ly\alpha})\overline{Q}_H(m)\\
&\simeq&
  4.51\times 10^{22}~{\rm erg~s^{-1}~Hz^{-1}}
 \left[\frac{\overline{Q}_H(m)}{10^{49}~{\rm s}^{-1}}\right]
\nu_{ly\alpha}\phi(\nu-\nu_{ly\alpha}),
  \label{eq:lya}
\end{eqnarray}
where $h\nu_{ly\alpha}=10.2~{\rm eV}$, $f_{ly\alpha}=0.64$,
and $\nu_{ly\alpha}=2465~{\rm THz}$.
Note that $f_{ly\alpha}=0.64$ was derived as follows: 
every hydrogen-ionizing photon results in a $n=2\rightarrow 1$
transition. 
(This is because every electron that goes directly to the ground state 
from $n \ge 3$ emits Lyman-series photons which are strongly absorbed, 
creating another excited atom, and this process repeats until all electrons 
end up in $n=2$ state.)  
About 2/3 of the time a $n=2\rightarrow 1$ transition creates a 
Lyman-$\alpha$ photon via
$2p\rightarrow 1s$ transition
and about 1/3 of the time it emits continuum emission via 2-photon decay
of $2s\rightarrow 1s$. The precise value of $f_{ly\alpha}$ depends slightly
on the temperature of gas, and for a gas at 20,000 K the value of $f_{ly\alpha}$ 
is 0.64 \citep{spitzer:1978}.
Finally, we ignore helium recombination lines 
as their flux is 
at most 6\% of the hydrogen-ionizing flux even for metal-free stars 
(see Table~1 and 4 of \citet{schaerer:2002}).
As for a line profile, we take it to be a delta function:
\begin{equation}
  \phi(\nu-\nu_{ly\alpha}) = \delta^D(\nu-\nu_{ly\alpha}).
\end{equation}
This is an excellent approximation as we are interested in
the background intensity which is integrated over a broad range
of redshifts. If we are, on the other hand, interested in a spectrum
of individual objects with fine spectral resolution,
more accurate calculations are required \citep{loeb/rybicki:1999}.
We have confirmed validity of our approximation by comparing the 
resulting spectrum with and without the exact line profile taken 
into account. 

It should be emphasized that the escape fraction, 
a fraction of ionizing photons escaping from nebula, does not alter luminosity
of Lyman-$\alpha$ very much.
This is because all of the ionizing photons will eventually be converted to 
Lyman-$\alpha$ photons which, in turn, will escape freely via the cosmological redshift.
Therefore, our prediction is free from uncertainty in the escape fraction.
In other words, we do not care where those Lyman-$\alpha$ photons come from
as far as the mean intensity is 
concerned\footnote{%
  However, the escape fraction should affect fluctuations as it changes 
morphology of the ionized region.}.
If most of the ionizing photons
escape from nebulae, Lyman-$\alpha$ photons should come from the 
intergalactic medium (but not too far away from nebulae;
otherwise, the Lyman-$\alpha$ signal would be spread over a large
frequency range and the signal would be suppressed). 
If none of the ionizing photons escape, Lyman-$\alpha$ photons should come 
from nebulae. In both cases, the resulting flux in Lyman-$\alpha$ should be
about the same.

\subsection{Two-photon Emission}

Luminosity of 2-photon emission is given by
\begin{eqnarray}
  \nonumber
  \overline{L}_\nu^{2\gamma}(m) 
  &=& \frac{2h\nu}{\nu_{ly\alpha}}(1-f_{ly\alpha})P(\nu/\nu_{ly\alpha})\overline{Q}_H(m)\\
  &\simeq& 4.24\times 10^{22}~{\rm erg~s^{-1}~Hz^{-1}}~\left[\frac{\overline{Q}_H(m)}{10^{49}~{\rm s}^{-1}}\right]
   \frac{\nu~P(\nu/\nu_{ly\alpha})}{\nu_{ly\alpha}/2},
   \label{eq:twophoton}
\end{eqnarray}
where $P(y)dy$ 
is the normalized probability per 2-photon decay of getting
one photon in the range $dy=d\nu/\nu_{ly\alpha}$. This formula is easily understood: 
again, every single ionizing photon results in a $n=2\rightarrow 1$ transition,
and $1-f_{ly\alpha}\simeq 1/3$ (more precisely 0.36 for $T_g=20,000$~K) 
of time it emits 2 photons via 2-photon decay.
(Therefore there is a factor of 2 multiplying $h\nu$.) 
We have fitted the data given in Table~4 of \citet{brown/mathews:1970}
to obtain\footnote{\citet{brown/mathews:1970} tabulate
$g(y)\equiv 2h y P(y)$, where $h=6.626\times 10^{-27}~{\rm erg~Hz^{-1}}$ is 
Planck's constant.}
\begin{equation}
  P(y) = 1.307 - 2.627(y-0.5)^2 + 2.563(y-0.5)^4 - 51.69(y-0.5)^6.
\end{equation}
Note that $P(y)$ is normalized such that $\int_0^1 P(y)dy =1$.
(This fitting formula gives $\int_0^1 P(y)dy = 1.0047$.)

\section{ENERGY BUDGET}
\label{sec:energy}

\subsection{Initial Mass Spectrum}
\label{sec:imf}

In order to calculate a typical spectrum of radiative efficiency, 
$\langle \epsilon_{\nu}\rangle$ (Eq.~[\ref{eq:avg}]), one needs
to specify the mass spectrum of stars, $f(m)$,
which determines the mean stellar mass of star formation. 
(For the precise definition of $f(m)$, see Appendix~\ref{sec:emissivity}.)
This is important because, depending on which mass 
is the most typical one, hardness of the emerging stellar spectrum
changes significantly. (Hardness affects the ratio of 
energy in Lyman-$\alpha$ to that in stellar continuum.)
Unfortunately, since we have very little knowledge of $f(m)$ for early stars, we are not able to estimate the proper error that would result in changing $f(m)$.
While we try to explore a range of models for $f(m)$, 
one should keep in mind that our exploration of the form of $f(m)$ is not exhaustive. 

We use three different mass spectra:
(a) Salpeter \citep{salpeter:1955}:
\begin{equation}
  f(m) \propto m^{-2.35},
  \label{eq:salpeter}
\end{equation}
and (b) Larson \citep{larson:1998}:
\begin{equation}
  f(m) \propto m^{-1}\left(1+\frac{m}{m_c}\right)^{-1.35},
  \label{eq:larson}
\end{equation}
which matches Salpeter's in the limit of 
$m_c\rightarrow 0$, and one can explore a variety of models
by changing one parameter, $m_c$. We shall assume $m_c=50~M_\sun$.
Finally, (c) a top-heavy spectrum:
\begin{equation}
  f(m) \propto \left\{ 
  \begin{array}{ll}
  m^{-1}, & 100<m<500~{\rm M_\sun}\\
  0, & {\rm otherwise} 
  \end{array}\right.
  \label{eq:topheavy}
\end{equation}
which might be possible 
for the primordial metal-free stars \citep{bromm/larson:2004}. 
(Note that $mf(m)$ is flat for $100<m<500~{\rm M_\sun}$.)
The normalizations are given by 
\begin{equation}
  \int_{m_1}^{m_2} dm~f(m) = 1.
\end{equation}
The choice of the mass range is somewhat arbitrary.
Throughout this paper, we shall assume $m_1=5~{\rm M_\sun}$ and $m_2=150~{\rm M_\sun}$
for the Salpeter and Larson mass spectra, whereas $m_1=100~{\rm M_\sun}$ and 
$m_2=500~{\rm M_\sun}$ for the top-heavy spectrum.
(We shall explain the reason for $m_1=5~{\rm M_\sun}$ later in \S~\ref{sec:comparison}.)
The mean stellar masses (Eq.~[\ref{eq:mstar}]) are 
13.6, 27.4, and 248.5 ${\rm M_\sun}$ for the Salpeter,
Larson, and top-heavy spectrum, respectively.

\subsection{Metallicity}
\label{sec:metallicity}

The next ingredients are the stellar luminosity-mass relation,
$\overline{L}^*_\nu(m)$, the ionizing flux-mass relation, $\overline{Q}_H(m)$,
the stellar lifetime-mass relation, $\tau(m)$, and the effective 
temperature-mass relation, $T_{eff}(m)$.
Since these relations mainly depend on metallicity, we explore
two cases: (1) metal-free ($Z=0$) stars, and 
(2) metal-poor ($Z=1/50Z_{\sun}$) stars. 


\begin{figure}
\centering \noindent
\includegraphics[width=18cm]{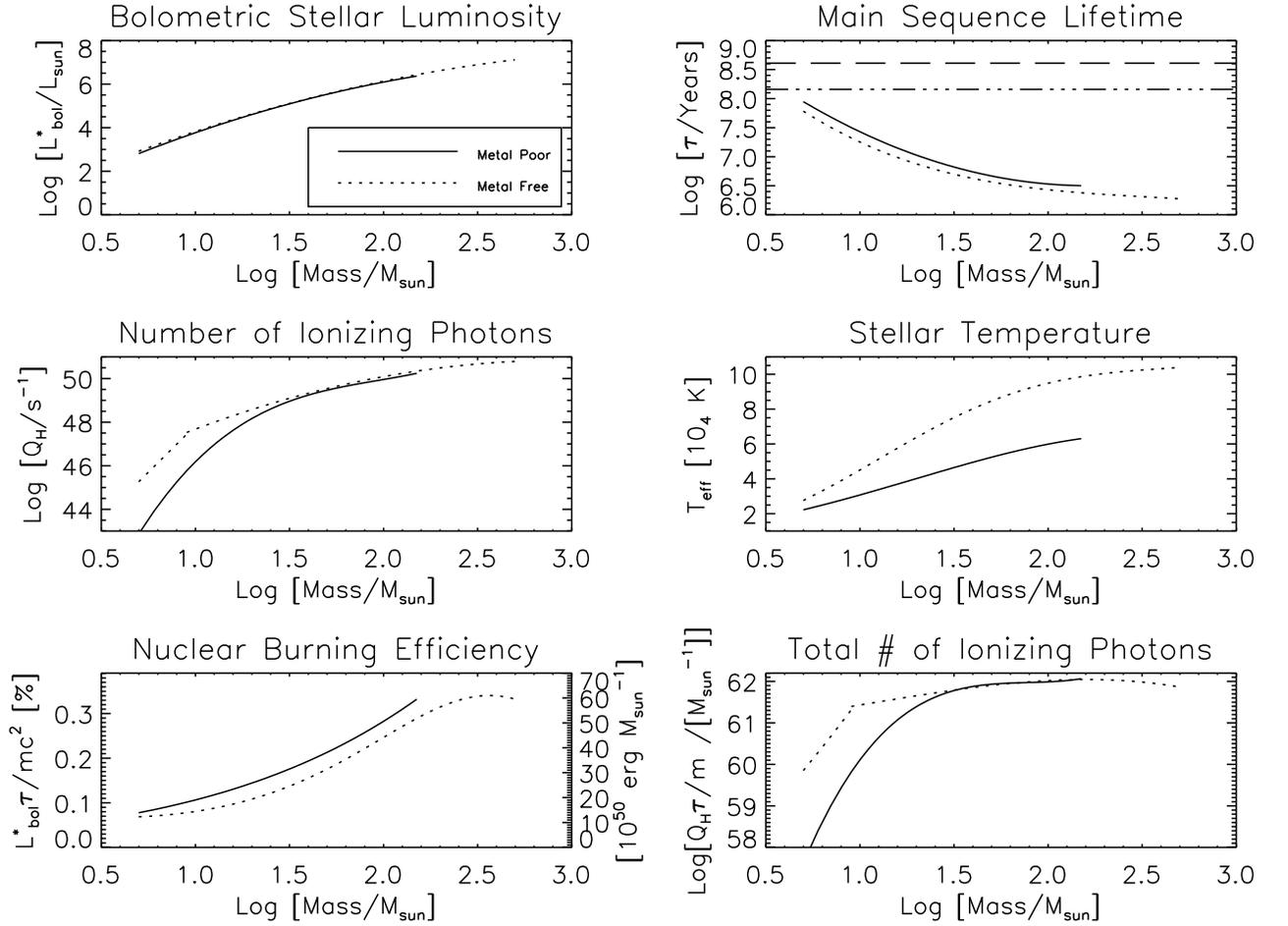}
\caption{%
  Stellar properties.
  ({\it Top-left}) The stellar bolometric luminosity, $L^*_{bol}$; 
  ({\it Top-right}) the stellar main-sequence lifetime, $\tau$; 
  ({\it Middle-left}) the number of hydrogen-ionizing photons emitted per unit time,
  $\overline{Q}_H$; 
  ({\it Middle-right}) the stellar effective temperature, $T_{eff}$;  
  ({\it Bottom-left}) nuclear burning efficiency, with the axis on the right showing bolometric 
  radiative energy per stellar mass over the main sequence lifetime; and 
  ({\it Bottom-right}) the total number of hydrogen-ionizing photons per unit stellar mass,
  for different stellar metallicities. 
  In the top-right panel, the dashed and dotted horizontal lines show the age of the 
  universe at $z = 7$ and 15, respectively.
}%
\label{fig:star}
\end{figure}


\subsubsection{Metal-free Stars}
\label{sec:metalfree}

\citet{schaerer:2002} has calculated emission properties
of metal-free stars. We use his fitting formulas for 
$\overline{Q}_H(m)$ and $\tau(m)$ 
over $m=5-500~M_\sun$ (Table~6 of \citet{schaerer:2002}):
\begin{eqnarray}
   \log_{10}\left[\overline{Q}_H/{\rm s^{-1}}\right]
      &=&\left\{
      \begin{array}{ll}
      43.61 + 4.90x - 0.83x^2 & 9-500~M_\sun,\\
      39.29 + 8.55x & 5-9~M_\sun,
      \end{array}\right.\\
   \log_{10}\left[\tau/{\rm yr}\right]
      &=& 9.785 - 3.759x + 1.413x^2
        - 0.186x^3,
\end{eqnarray}
where $x\equiv \log_{10}(m/M_\sun)$.
We calculate the stellar radius as
\begin{equation}
  \label{eq:radius}
  4\pi R^2_*(m) = \frac{L_{bol}^*(m)}{\sigma T_{eff}^4(m)}, 
\end{equation}
where $\sigma=5.67\times 10^{-5}~{\rm erg~s^{-1}~cm^{-2}~K^{-4}}$ 
is the Stephan-Boltzmann constant. 
The bolometric stellar luminosity before absorption, $L^*_{bol}$, and the effective 
temperature, $T_{eff}$, are given in Table~3 of \citet{schaerer:2002}.  
These were used to obtain 
fitting formulas for $T_{eff}$ and $L^*_{bol}$, which are good for masses anywhere 
from $5-1000 ~{\rm M_\sun}$:
\begin{eqnarray}
  \log_{10}\left[L_{bol}^*/L_\sun\right]&=& 0.4568+3.897x-0.5297x^2\\
  \log_{10}\left[T_{eff}/{\rm K}\right]&=& 3.639+1.501x-0.5561x^2+0.07005x^3
\end{eqnarray}
Note that $L_{bol}^*$ and $T_{eff}$ were 
calculated for the zero-age main sequence
stars, whereas $\overline{Q}_H$ has been averaged over the main-sequence lifetime.
Strictly speaking, the former quantities should have also 
been averaged over the stellar lifetime; however, we shall ignore such 
a subtlety and use the zero-age values as tabulated in \citet{schaerer:2002}.

\subsubsection{Metal-poor Stars}
\label{sec:metalpoor}

For stars with $Z=1/50Z_{\sun}$, we use the fitting formula for 
$\overline{Q}_H(m)$ and  $\tau(m)$ given in Table~6 of \citet{schaerer:2002}:
\begin{eqnarray}
   \log_{10}\left[\overline{Q}_H/{\rm s^{-1}}\right]
      &=& 27.80 + 30.68x - 14.80x^2
        + 2.50x^3,\\
   \log_{10}\left[\tau/{\rm yr}\right]
      &=& 9.59-2.79x+0.63x^2,
\end{eqnarray}
where $x\equiv \log_{10}(m/M_\sun)$.  The formula for stellar lifetimes 
and ionizing photons is good from $7-150~{\rm M_\sun}$; 
we shall extrapolate it down to $5~{\rm M_\sun}$.  
We calculate the stellar radius (Eq.~[\ref{eq:radius}]) using
$L^*_{bol}(m)$ and $T_{eff}(m)$.  This was fit from stellar models given in \citet{lejeune:2001}.
The fitting formulas were thus obtained as
\begin{eqnarray}
  \log_{10}[L^*_{bol}/L_\sun]&=& 0.138+4.28x-0.653x^2,\\
  \log_{10}[T_{eff}/{\rm K}]&=&   3.92+0.704x-0.138x^2.
\end{eqnarray}
Again, $L_{bol}^*$ and $T_{eff}$ were calculated for the zero-age main sequence
stars, whereas $\overline{Q}_H$ has been averaged over the main-sequence lifetime.

\subsubsection{Stellar Properties}
\label{sec:stellarprop}

Figure~\ref{fig:star} shows 
the bolometric stellar luminosity before absorption, $L^*_{bol}$ (top-left panel), 
the main-sequence lifetime, $\tau$ (top-right), 
the number of hydrogen-ionizing photons per second, $\overline{Q}_H$ (middle-left), and
the stellar effective temperature, $T_{eff}$ (middle-right),
for $Z=0$ (labeled as ``metal-free'') and $Z=1/50Z_{\sun}$ (``metal-poor'').
The bolometric luminosity is very similar for metal-free and metal-poor stars
at the same stellar mass down to $\sim 10~{\rm M_\sun}$ \citep{tumlinson/shull:2000},
and is almost identical for more massive stars ($\gtrsim 100~{\rm M_\sun}$)
\citep{bromm/kudritzki/loeb:2001, abel/bryan/norman:2002}.
Since metal-free stars had to begin their nuclear burning via the p-p chain,
which is less efficient than the CNO cycle because of weak interactions, 
the temperature of metal-free stars must be maintained higher than
that of metal-poor stars to prevent gravitational collapse
\citep{tumlinson/shull:2000}. Since the luminosity is similar, this property
makes the size of metal-free stars smaller and the main-sequence 
lifetime slightly shorter than those of metal-poor stars.
On the other hand, metal-free stars produce more hydrogen-ionizing photons
than metal-poor stars, particularly for $m\lesssim 30~M_\sun$,
owing to their higher temperature (the spectrum is harder).

The bottom panels of Figure~\ref{fig:star} show 
quantities more relevant to the radiative efficiency, $\epsilon_{\nu}(m)$. 
The first panel shows the ratio of the stellar bolometric energy to the rest mass 
energy.
This figure shows that for $m\simeq 10-100~M_\sun$ anywhere from $0.1$ to 
$0.3\%$ of 
the rest mass energy of the star goes into radiative energy via nuclear 
fusion; thus,
this quantity represent a ``nuclear burning efficiency'' of stars.   
The metal-poor stars radiate slightly more energy over their lifetime than
the metal-free stars, as they live slightly longer and the bolometric luminosity 
is about the same.
On the right, the total number of ionizing photons per unit stellar mass, 
$\overline{Q}_H\tau/m$, is shown. The metal-poor stars emit significantly
less ionizing photons for $m\lesssim 30~M_\sun$: this property becomes 
important when we interpret the predicted spectrum of the near infrared background.

\subsection{Energy Spectrum}
\label{sec:epsilon}


\begin{figure}
\centering \noindent
\includegraphics[width=16cm]{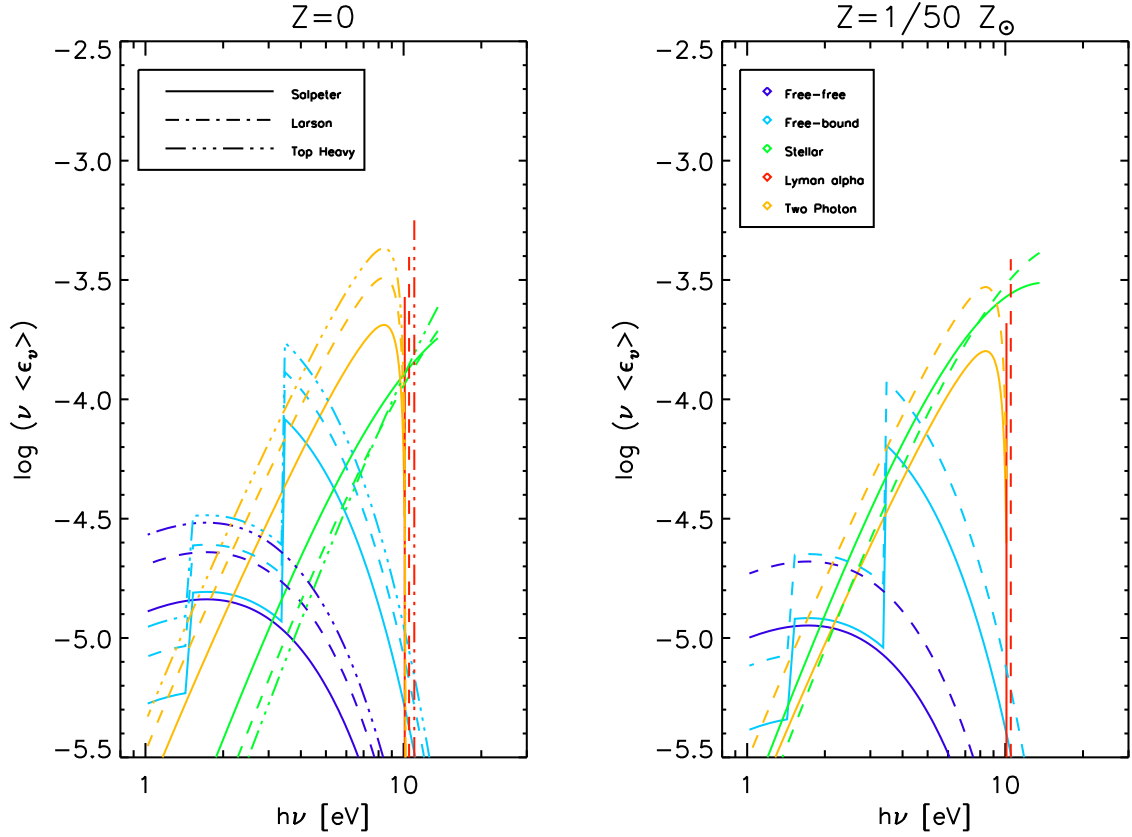}
\caption{%
Spectrum of radiative efficiency, $\nu\langle\epsilon_{\nu}\rangle$, 
for different metallicities: $Z=0$ (left) and $Z=1/50Z_{\sun}$ (right). 
The radiative efficiency is defined as a fraction
of the stellar rest mass energy that goes into free-free (the lowest purple lines), 
free-bound (the ``saw-shaped'' cyan lines), stellar black-body (green), two-photon emission (orange) or
Lyman-$\alpha$ line (the vertical red lines) emission.
The solid, dot-dashed, and triple-dot-dashed lines show the Salpeter
(Eq.~[\ref{eq:salpeter}]), Larson (Eq.~[\ref{eq:larson}]),
and top-heavy (Eq.~[\ref{eq:topheavy}]) initial mass spectra, respectively.
}%
\label{epsilon}
\end{figure}


Using these fitting formulas and the initial mass spectra, we calculate 
a spectrum of radiative efficiency averaged over the mass spectrum
for $m=m_1-m_2$ (Eq.~[\ref{eq:avg}]).
Figure~\ref{epsilon} shows $\nu\langle \epsilon_{\nu}^{\alpha}\rangle$
for the stellar (Eq.~[\ref{eq:stellar}]),
nebular continuum (free-free and free-bound) (Eq.~[\ref{eq:ffbf}]),  
Lyman-$\alpha$ (Eq.~[\ref{eq:lya}]), and two-photon (Eq.~[\ref{eq:twophoton}]) 
emission. The nebular continuum dominates at low energy, 
while the stellar, Lyman-$\alpha$ and two-photon emission dominate at high energy, 
as expected from their spectral shape.
Since metallicity changes hardness of the stellar spectrum,
it affects the ratio of energy in Lyman-$\alpha$ and two-photon emission to stellar
emission energy: the harder the spectrum is, the more the ionizing
photons are emitted, and thus the more the Lyman-$\alpha$ and two-photon emission
emerge\footnote{However, this is not always the case. 
The bottom-right panel of Figure~\ref{fig:star} shows that metal-poor stars actually
emit as many ionizing photons per stellar mass as metal-free stars
for $m\gtrsim 30~M_\sun$; thus, if the mean stellar mass of 
metal-poor stars were $\gg 30~M_\sun$, metal-poor stars would result 
in as many Lyman-$\alpha$ photons as metal-free stars.}.
This explains why the metal-free stars have
much more energy in Lyman-$\alpha$ and two-photon emission than in 
stellar emission. On the other hand, the metal-poor stars have more energy in 
stellar emission.
For the same reason, heavier mass spectra tend to produce 
more energy in Lyman-$\alpha$ and two-photon emission than in stellar emission. 
In both cases, however, the total radiative efficiency is about 
the same: $\nu\langle \epsilon_{\nu}\rangle\sim 10^{-3}$.
This is merely an approximate conservation of energy:
initially all the energy was generated by nuclear burning
in stars. 
The generated energy is then radiated or reprocessed, but the 
sum should be more or less the same as the input energy.
(Of course conservation cannot be exact because we have ignored
other emission processes such as Balmer lines, helium or metal lines,
etc. If the HII region expands, additional energy would be lost to
expansion.)
This property makes the prediction of the near infrared background
very robust, up to an unknown star formation rate, $\dot{\rho}_*$,
which will be constrained by a comparison to the observational data.

\section{SPECTRUM OF THE NEAR INFRARED BACKGROUND}
\label{sec:irb}

\subsection{Dependence on Metallicity and Initial Mass Spectrum}
\label{sec:metalimf}

\begin{figure}
\centering \noindent
\includegraphics[width=16cm]{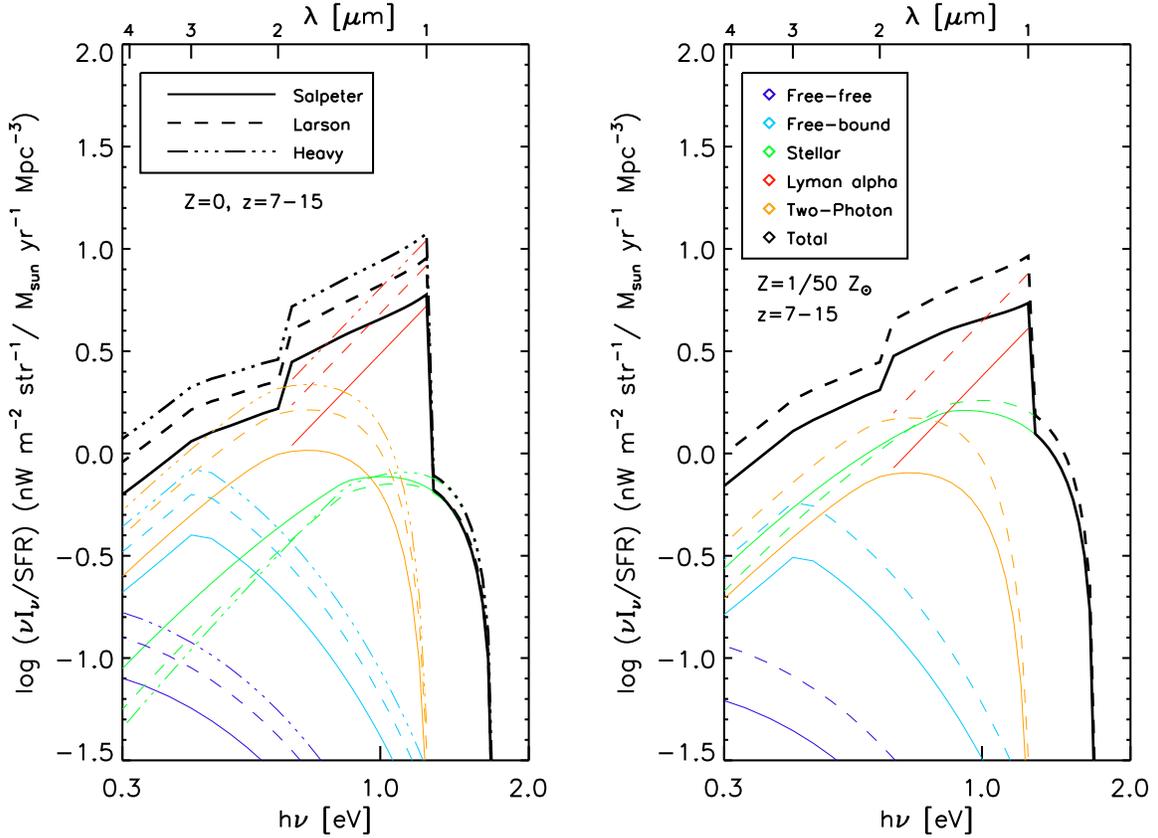}
\caption{%
  Spectrum of the near infrared background, $\nu I_{\nu}/\dot{\rho}_*$, 
  for star formation from $z=7-15$ with various assumptions about metallicity
  and initial mass spectrum. (``SFR'' in the axis label denotes $\dot{\rho}_*$.) 
  The left panel shows the metal-free case ($Z=0$), 
  while the right panel shows the metal-poor case ($Z=1/50Z_{\sun}$). The solid, dashed, and
  triple-dotted-dashed lines represent the Salpeter, Larson, and top-heavy mass 
  spectrum, respectively. The thick black lines show the total spectrum, while 
  the thin purple, cyan, green, red and orange lines show individual contribution from 
  free-free, free-bound, stellar, Lyman-$\alpha$ and two-photon emission, respectively.
}%
\label{fig:nirb}
\end{figure}

\begin{figure}
\centering \noindent

\includegraphics[width=14cm]{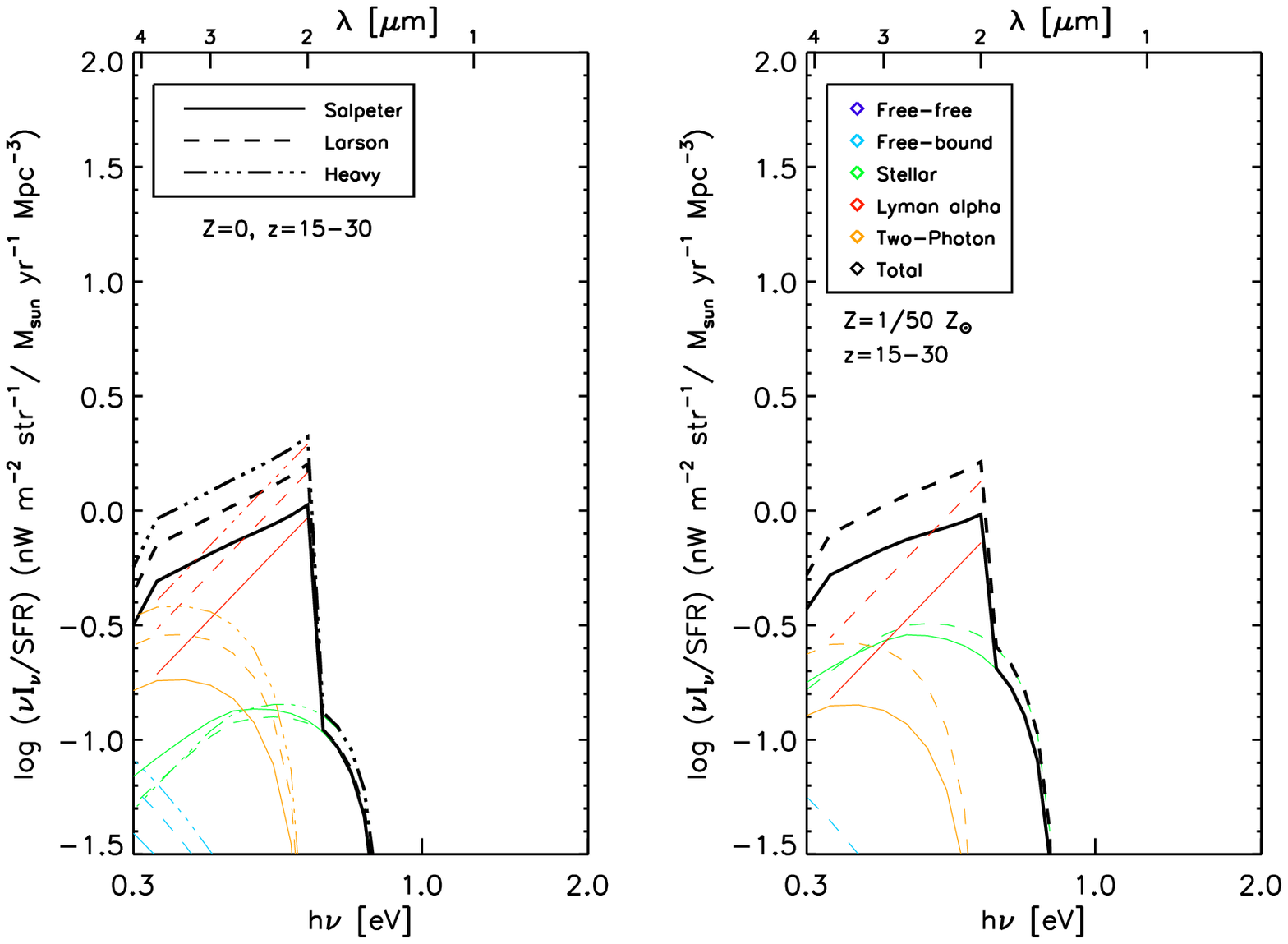}
\includegraphics[width=14cm]{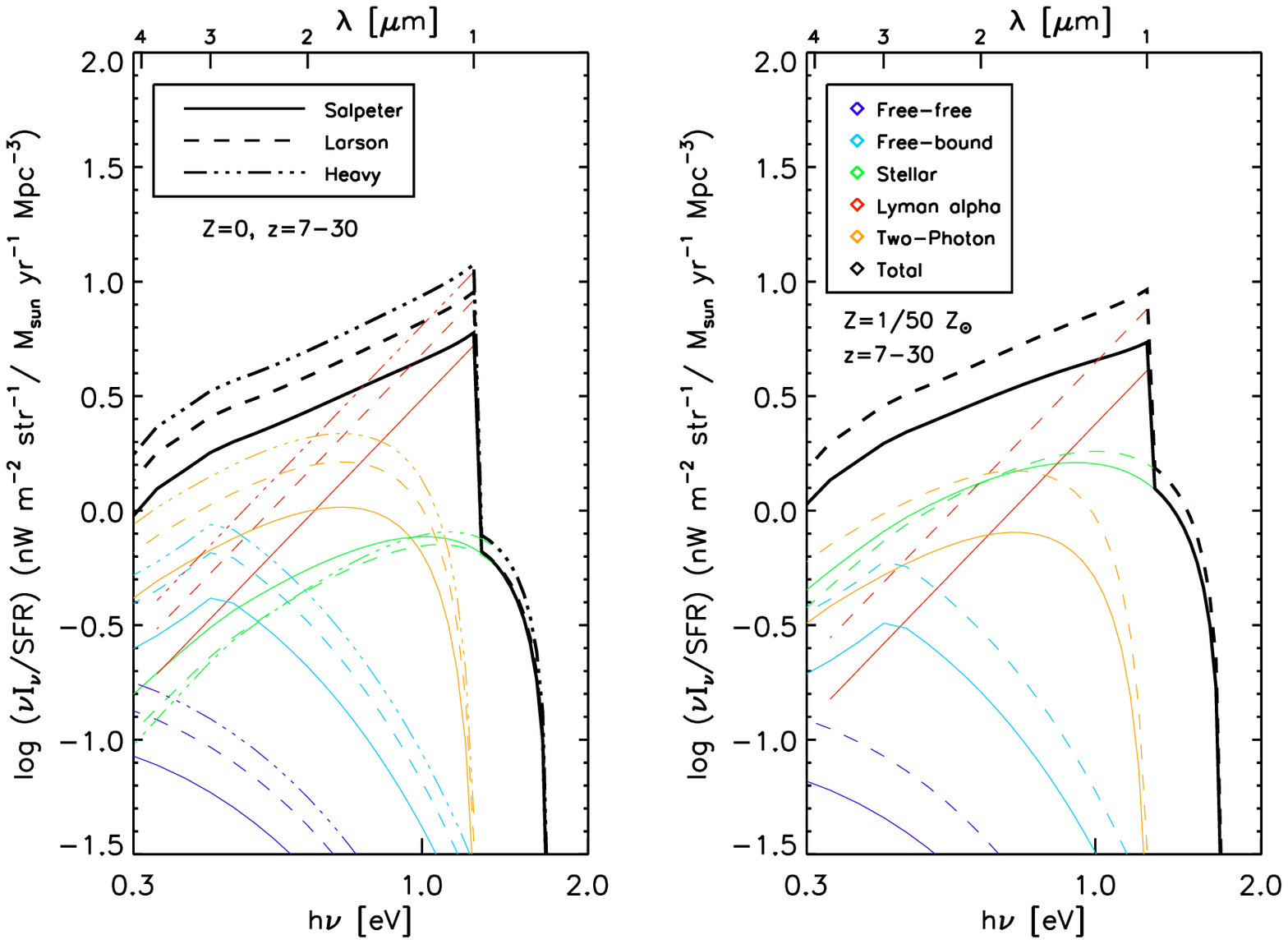}
\caption{%
   The same as Figure~\ref{fig:nirb} for $z=15-30$ ({\it top}) and 
   $z=7-30$ ({\it bottom}).  
}%
\label{fig:nirb2}
\end{figure}

By integrating the volume emissivity over redshift,
we obtain the background intensity spectrum of the near infrared from
early stars (Eq.~[\ref{eq-inu-order}]).
To do this, however, one needs to specify the evolution of star formation 
rate over time, $\dot{\rho}_*(z)$, which is unknown. 
Therefore, for simplicity, we shall assume
that the star formation rate is constant over time, at least for the redshift range of 
interest. In other words, we calculate the intensity spectrum for a given ``time-averaged''
star formation rate.  
Figure~\ref{fig:nirb} and \ref{fig:nirb2} show 
$\nu I_\nu/\dot{\rho}_*$ for stars in three redshift ranges, $z=7-15$, $15-30$, and $7-30$.
These figures clearly show that the intensity at $1-2~\mu{\rm m}$
is almost entirely determined by the contribution at $z=7-15$. (Note that Lyman-$\alpha$ lines
at $z=7-15$ are redshifted to $1-2~\mu{\rm m}$.)
Therefore, the spectrum of the near infrared background at $1-2~\mu{\rm m}$
constrains the star formation rate at $z=7-15$!

Table~\ref{tab:nuInu} summarizes values of $\nu I_\nu/\dot{\rho}_*$ averaged
over $1-2~\mu{\rm m}$.
Within $1-2~\mu{\rm m}$, the intensity is dominated by Lyman-$\alpha$
emission.  For metal poor stars, there is also a significant contribution 
from stars themselves, 
which brings the overall intensity for metal poor and metal free stars to be about the same. 
This seems striking, but is merely a consequence of an approximate
energy conservation, as discussed in \S~\ref{sec:energy}\footnote{We thank Paul R. Shapiro
for pointing out potential importance of metal-poor stars for 
the near infrared background.}. 
Therefore, the predicted intensity is not sensitive to stellar metallicity.

\begin{figure}[h!tb]
\centering \noindent
\includegraphics[width=12cm]{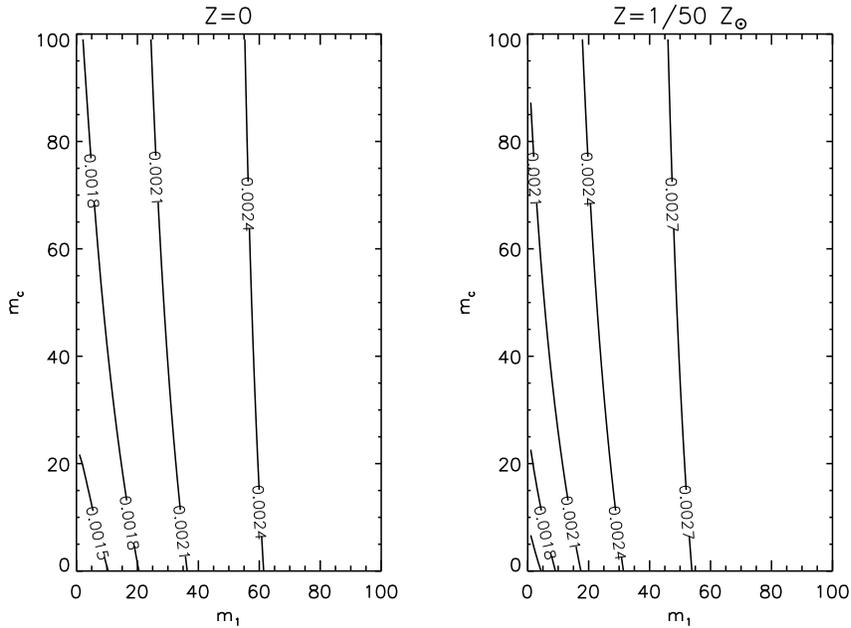}
\caption{%
Mass-weighted mean nuclear burning efficiency, $\langle L_{bol}^*\tau/(mc^2)\rangle$,
 resulting from varying the initial mass spectrum.
Each contour shows a different nuclear burning efficiency, and
$m_1$ and $m_c$ are the lower mass limit of star formation and the critical mass
in Larson's mass spectrum (Eq.~[\ref{eq:larson}]), respectively.
}%
 \label{fig:contour}
\end{figure}

As for dependence on the initial mass spectrum, $f(m)$, heavier mass spectra tend to
give higher background intensities.
Energetics implies that dependence of $\nu I_\nu/\dot{\rho}_*$ on metallicity or $f(m)$
should be essentially given by that of the nuclear burning efficiency averaged over a population of stars.  
The last column of Table~\ref{tab:nuInu} shows the mass-weighted mean 
nuclear burning efficiency, 
$\langle L_{bol}^*\tau/(mc^2)\rangle$,
which is tightly correlated with the total signal. Therefore, one can explore dependence of 
the near infrared background on these parameters by simply calculating the nuclear burning efficiency dependence
on these parameters.
In order to illustrate how nuclear burning efficiency changes with respect to the shape of $f(m)$,
we show the efficiency for various values of the lower mass limit, $m_1$, and the critical mass, $m_c$, for the 
Salpeter (Eq.~[\ref{eq:salpeter}]) and Larson (Eq.~[\ref{eq:larson}]) initial mass spectra in Figure~\ref{fig:contour}.
The average nuclear burning efficiency for $m_1>20~M_\sun$ depends very weakly on $m_c$, while 
the dependence is stronger for $m_1<20~M_\sun$. 
Dependence on $m_1$ also becomes stronger as $m_1$ becomes smaller.
Overall, for Larson's mass spectrum, different $m_1$ and $m_c$ may change the predicted
intensity by a factor of a few, but not much more. 
However, one should keep in mind that other forms of $f(m)$ that we have not explored here 
might change the predicted intensity by a larger factor.

Mass loss from stars can also affect the total spectrum.  \citet{schaerer:2002} gives properties of metal free stars undergoing strong mass loss.  Using these values, nuclear burning efficiency would be increased by a factor of $1.25$.  The strong mass loss, if any, therefore reduces the inferred star formation rate by roughly 25-50 \%.  (\S~\ref{sec:sfr1})


\begin{table}[t]
\begin{center}
\begin{tabular}{|l|l||l|l|l|l|l|l||l|}
\hline
Metallicity& IMF       &  Stellar & Free-free & Free-bound & Ly-$\alpha$ 
&2-photon& Total &$10^{3}\langle\frac{L^*_{bol}\tau}{mc^2}\rangle$ \\
\hline
Metal-Poor & Salpeter  & 1.46 &0.00761&0.0590  & 2.10&0.530 &4.16 & 1.35 \\
           & Larson    & 1.60 &0.0141&0.109    & 3.90 &0.984 &6.61&1.84  \\
Metal-Free & Salpeter & 0.678 & 0.00979&0.0759  & 2.71&0.683 &4.15&1.25  \\
           & Larson    & 0.605 &0.0154&0.120  & 4.27&1.08 &6.08 &1.73    \\
           & Top Heavy & 0.663 & 0.0205&0.159 & 5.67&1.43 &7.95 &3.23    \\
\hline
\end{tabular}
\caption{%
   Values of $\nu I_{\nu}/\dot{\rho}_*$ averaged over $1-2~{\mu\rm m}$
   in units of ${\rm nW~m^{-2}~sr^{-1}/M_{\odot}~yr^{-1}~Mpc^{-3}}$ from 
   each radiative process. 
   Here $\dot{\rho}_*$ is the mean star formation rate for $m>5~M_\sun$ at $z=7-15$,
   and $\langle L^*_{bol}\tau/(mc^2)\rangle$ is the mass-weighted
  mean nuclear burning efficiency of stars. 
}%
\label{tab:nuInu}
\end{center}
\end{table}

%


\subsection{Comparison with Previous Work}
\label{sec:comparison}

\citet{santos/bromm/kamionkowski:2002} calculated the near infrared
background from metal-free stars, assuming that all early stars 
contributing to the background light have $m=1000~M_\sun$
(i.e., $f(m)=\delta^D(m-1000~M_\sun)$). While the stellar,
Lyman-$\alpha$, and free-free emission were included in their calculations,
the two-photon and free-bound emission were ignored.
As we have shown in Figure~\ref{fig:nirb}, the contribution from 
free-bound transition to $n=2$ is much larger
than the free-free contribution, and the contribution from two-photon emission is
 as important as that from stellar emission; thus, they must be included.

\citet{salvaterra/ferrara:2003} assumed metal-free stars but explored different
initial mass spectra, varying $m_c$ of Larson's spectrum.
There is a subtle but important difference between
their approach and our approach. When they modeled the volume emissivity, 
they did not allow for the faster rate of death of higher mass stars.
In other words, their formula for the volume emissivity implicitly assumed that 
stars lived longer than the Hubble time. We find that the volume emissivity 
that is not corrected for dead stars (Eq.~[\ref{eq:psf}]) agrees with their formula 
(Eq.~[5] and [12] in \citet{salvaterra/ferrara:2003}). This assumption leads to an 
overestimation of contributions from higher mass stars which live much shorter than 
the Hubble time.
 The physical reason is because higher mass stars live shorter and die sooner, and 
 therefore there are fewer massive stars around to emit energy at a given $z$.
 By properly taking into account the faster rate of death of stars, we have obtained
 the correct formula (Eq.~[\ref{eq:p}]) which explicitly depends
 on the stellar main-sequence lifetime.
On the other hand, if stars live longer than or comparable to the Hubble time, then 
our approximation breaks down and one should do the integral in Eq.~[\ref{eq:emformula}]. 
This is precisely why we have restricted our attention only to fairly massive stars,
$m>5~M_\odot$, for which the age of the universe is definitely longer than the 
main-sequence lifetime at redshifts of interest.

\section{IMPLICATIONS FOR THE COSMIC STAR FORMATION RATE}
\label{sec:sfr}

\subsection{Star Formation Rate at $z=7-15$}
\label{sec:sfr1}

Comparing the predicted values of $\nu I_\nu/\dot{\rho}_*$ (Table~\ref{tab:nuInu}) to the 
measured data, 
we can constrain the star formation rate $\dot{\rho}_*$. 
The near infrared background has been determined with various satellites, 
such as the Diffuse Infrared Background Experiment (DIRBE) on the Cosmic Background Explorer 
\citep{hauser/etal:1998, boggess/etal:1992} and 
the Near Infrared Spectrometer (NIRS) on the InfraRed Telescope in Space (IRTS) 
\citep{matsumoto/etal:2005}.  
Table~\ref{tab:obs} summarizes the observational data.
A significant uncertainty exists in the observational data, largely because 
of uncertainty in subtraction of the zodiacal emission. 
A large difference between \citet{wright:2001} and \citet{cambresy/etal:2001},
which have used the same data (DIRBE), is entirely due to difference in the
zodiacal light models. 
One may summarize the current measurement of the cosmic near infrared background
as $\sim 2-50~{\rm nW~m^{-2}~sr^{-1}}$
in 1--2~$\mu$m, which includes the 1-$\sigma$ lower bound of the lowest measurement and 
the 1-$\sigma$ upper bound of the highest measurement.  \footnote{A recent analysis of blazar spectrum by \citet{aharonian/etal:2005} suggested that the intensity of the near infrared background may be lower, with an upper limit of $\sim (14 \pm 4)~{\rm nW~m^{-2}~sr^{-1}}$ from 1--2~$\mu$m, which is more consistent with the analysis of the DIRBE data by \citet{wright:2001}.  On the other hand, a recent detection of the fluctuations in the near infrared background may imply that at least some of the NIRB is from early stars \citep{kashlinsky/etal:2005}}.
Taking into account a scatter in theoretical predictions due to different 
assumptions about
metallicity and initial mass spectrum 
(see Table~\ref{tab:nuInu} and \ref{tab:generation}), we obtain
$\dot{\rho}_*\sim 0.3-12~{\rm M_{\odot}~yr^{-1}~Mpc^{-3}}$
at $z=7-15$. (Note that the error bar is not dominated by theory but 
by observational errors.)
What does this imply?


\begin{table}[t]
\centering
\begin{tabular}{|l|l||l|l|l||l|}
\hline
Instrument & $\nu I_\nu$   & Reference \\
\hline
DIRBE	& $14.6\pm 12.4$ & \citet{wright:2001} \\
			& $30.9\pm 12.8$ & \citet{cambresy/etal:2001} \\
NIRS	& $41.6\pm 9.7$ & \citet{matsumoto/etal:2005} \\ 
\hline
\end{tabular}
\caption{%
  Observational data of $\nu I_{\nu}$ averaged over $1-2~{\mu\rm m}$
in units of ${\rm nW~m^{-2}~sr^{-1}}$. Note that zodiacal emission, emission
of foreground stars, and emission from galaxies have been subtracted.
For DIRBE, the straight average of J- and K-band data (minus
$10~{\rm nW~m^{-2}~sr^{-1}}$ which is contributed by galaxies)
is quoted, and the error is estimated as $\sqrt{(\sigma^2_J+\sigma^2_K)/2}$.
For NIRS, the straight average of 9 measurements from 1.43 to 2.24~$\mu$m is quoted,
and the error is estimated similarly as $\sqrt{\sum_{i=1}^9\sigma_i^2/9}$.
Statistical errors are negligible compared to the systematic errors
in subtraction of the zodiacal emission. 
For details, see \citet{matsumoto/etal:2005}.}
\label{tab:obs}
\end{table}


\subsection{Stellar Mass Density Confronts Cosmic Baryon Density}
\label{sec:sfr2}

One must always make sure that the stellar mass density inferred
from star formation rate does not exceed the cosmic mean baryon density. 
Using the formula derived in Appendix~\ref{sec:omegastar}, we obtain the ratio
of {\it cumulative} mass density (which is not corrected for dead stars)
of stars formed at $7<z<15$ to the cosmic mean baryon density as
\begin{equation}
 \frac{\rho^{cum}_*(7<z<15)}{\rho_{b0}}
 \simeq
  0.0747~\left(\frac{0.024}{\Omega_bh^2}\right)
  \left(\frac{0.14}{\Omega_mh^2}\right)^{1/2}
  \left[\frac{\dot{\rho}_*(7<z<15)}{\rm M_\sun~yr^{-1}~Mpc^{-3}}\right],
  \label{eq:omegastar}
\end{equation}
where $\rho_{b0}=2.775\times 10^{11}(\Omega_bh^2)~{\rm M_\sun~Mpc^{-3}}$ is the 
present-day mean baryon density.
(Note that 
$\rho_*$ denotes comoving mass density.)
It follows from this equation that the inferred lower limit to the star 
formation rate,
$\dot{\rho}_*>0.3~{\rm M_{\odot}~yr^{-1}~Mpc^{-3}}$, requires
that more than 2\% of baryons in the universe should have been converted into stars. 
\citet{madau/silk:2005} argue that ``this is energetically and astrophysically daunting''.
It would be daunting, if the stars that were responsible for producing the near infrared 
background
lived longer than the age of the universe, and more than 2\% of baryons had {\it remained} 
locked up in the stars.
However, it is certain that stars lived much shorter than the age of the universe 
(see the top-right
panel of Figure~\ref{fig:star}), and the stellar mass density must be corrected
for dead stars; thus, the actual amount of baryons locked up in stars 
at any given time
between $z=7$ and 15 should be less than that is given by Eq.~[\ref{eq:omegastar}]. 
We derive the stellar mass density corrected for dead stars in Appendix~\ref{sec:density_dead},
which shows that the correct answer should lie between Eq.~[\ref{eq:omegastar}] and 
Eq.~[\ref{eq:omegastar}] divided by
 the mean number of generations of 
stars,
$N_{gen}$, given by
\begin{equation}
  N_{gen} = \frac{t(7<z<15)}{\int_{m_1}^{m_2}dm~f(m)\tau(m)},
  \label{eq:generation}
\end{equation}
and $t(7<z<15)=266$~million years is the cosmic time elapsed during $z=7-15$.
Table~\ref{tab:generation} tabulates $N_{gen}$ for various assumptions about 
metallicity and initial mass spectrum.
From this we conclude that, to explain the cosmic near infrared background 
by early generations
of stars, 0.016--12\% of baryons need to be 
processed in stars at a given time between $z=7$ and 15.  
If we take the lower 1-$\sigma$ limit, only 0.016--0.49\% of baryons
need to be processed in stars (depending on metallicity and mass spectrum);  
this is not a daunting requirement and does not exclude the stellar origin of
the cosmic near infrared background\footnote{Our argument so far has implicitly 
assumed that all of baryonic gas in the previous generation of stars is returned to the 
intergalactic medium and recycled in the subsequent generation of stars.
In reality, however, only a fraction of gas would be returned (and the rest of
gas would be locked up in compact remnants such as black holes); thus, 
the real requirement would be somewhat larger than 0.016--0.49\%.}.


\begin{table}[t]
\begin{center}
\begin{tabular}{|l|l|l|l|l|l|l|}
\hline
Metallicity& IMF       & $\dot{\rho}_*$ (Lower   & $\dot{\rho}_*$ (Upper & $N_{gen}$   & 
$\rho_*/\rho_b$ [\%] (Lower  &$\rho_*/\rho_b$ [\%] (Upper  \\
&&$1-\sigma$ limit)&$1-\sigma$ limit)&&$1-\sigma$ limit)&$1-\sigma$ limit)\\
\hline
Metal-Poor & Salpeter  & 0.48&12 & 7.3 & 0.49 &12.3\\
           & Larson    &0.30 &7.6 & 12.8 & 0.18 &4.41 \\
Metal-Free & Salpeter  & 0.48&12 & 10.0 & 0.36 &9.0 \\
           & Larson    & 0.33&8.2 & 16.7 & 0.15&3.7 \\
           & Top Heavy & 0.25&6.3 & 120  & 0.016 &0.39\\
\hline
\end{tabular}
\caption{%
  Implications of the current lower and upper $1-\sigma$ limits to the cosmic near infrared 
  background data, $\nu I_\nu=2~{\rm nW~m^{-2}~sr^{-1}}$ and $\nu I_\nu=50~{\rm nW~m^{-2}~sr^{-1}}$ , for 
  formation of early stars with $m>5~M_\sun$ during $z=7-15$. 
  The third and forth columns show the lower and upper limits, respectively, to
  the star formation rate in units of $M_\sun~{\rm yr^{-1}~Mpc^{-3}}$,
 the fifth column shows the mean number of 
  generations of stars (Eq.~[\ref{eq:generation}]), and  
  the sixth and seventh columns shows the lower and upper limits, respectively, to a ratio of stellar density
  to the mean cosmic baryon density in percentage.  
}%
\label{tab:generation}
\end{center}
\end{table}


\subsection{Comparison with Low Redshift Data}
\label{sec:lowz}

\begin{figure}[h!tb]
\centering \noindent
\includegraphics[width=12cm]{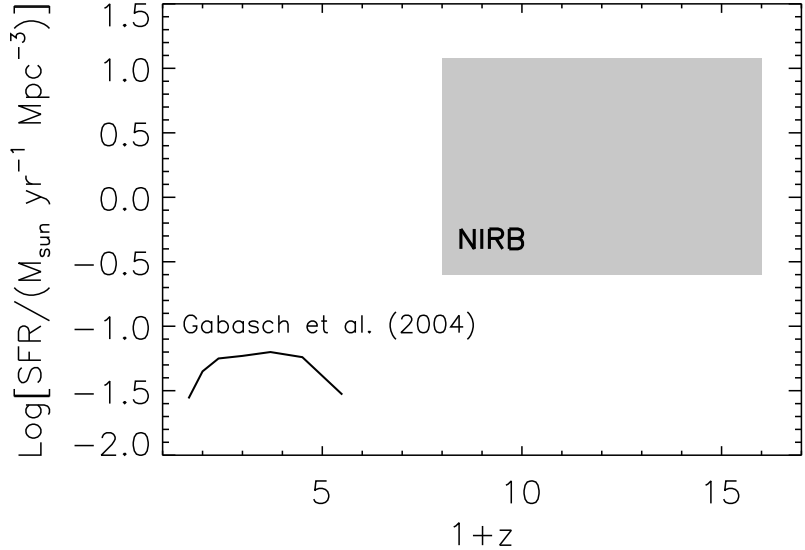}
\caption{%
Cosmic star formation rate.
The shaded region shows the star formation rate for $m>5~M_\sun$ constrained
by the current data of the cosmic near infrared background, 
$\nu I_\nu\sim 2-50~{\rm nW~m^{-2}~sr^{-1}}$, in 1--2~$\mu$m. 
The solid line shows the star formation rate at $z<5$
\citep{gabasch/etal:2004}. Note that the shaded region should be taken as a lower
bound. (See discussion in \S~\ref{sec:lowz}).
}%
\label{fig:sfr}
\end{figure}

How does the inferred star formation rate at $z=7-15$ compare to the low-$z$
rate? 
Figure~\ref{fig:sfr} compares the cosmic star formation rate at $z<5$ 
\citep{gabasch/etal:2004}\footnote{The rate at $z<5$ has been shifted upward by 0.35~dex 
to correct for dust extinction. More recent determination of the star formation rate  
by \citet{drory/etal:2005} agrees very well with \citet{gabasch/etal:2004}.}
 to that constrained by the near infrared background.
While uncertainty due to subtraction of the zodiacal light is large, it is quite
clear that the star formation rate  at $z=7-15$ 
required to account for the cosmic near infrared background data is much higher 
than that at $z<5$ by more than an order of magnitude.

It must be emphasized, however, that Figure~\ref{fig:sfr} is potentially misleading:
as we have already discussed, the star formation rate inferred from the near infrared
background is only for stars more massive than $5~M_\sun$. On the other hand,
the low-$z$ data are primarily dominated by low mass stars; thus, 
Figure~\ref{fig:sfr} might be comparing apples and oranges.
As such low mass stars do not contribute to the near infrared background,
it is not possible to infer their formation rate directly. One may still estimate
it by extrapolating the initial mass spectrum down to lower masses, and
by doing so the total star formation rate at $z=7-15$ should rise.
In other words, the constraint shown in Figure~\ref{fig:sfr} should be taken as a 
lower bound. Also, dust extinction (which we have ignored), if any, would make
the required star formation rate rise even higher. 


\section{METALLICITY CONSTRAINTS ON STAR FORMATION}
\label{sec:metals}

One of the ways to constrain early star formation is to take into account the amount of metals that can be produced without over-polluting the universe.  Metals ejected from a star have two origins: 1. stellar winds, which inject metals into the IGM over the course of the star's lifetime; and 2. the final disruption of the star.  Stars of low metallicity end their lives in different ways and produce different amounts of metals, according to the initial mass of the star \citep{heger/etal:2003, portinari/chiosi/bressan:1998, siess/livio/lattanzio:2002}.

\begin{figure}
\centering \noindent
\includegraphics[width=16cm]{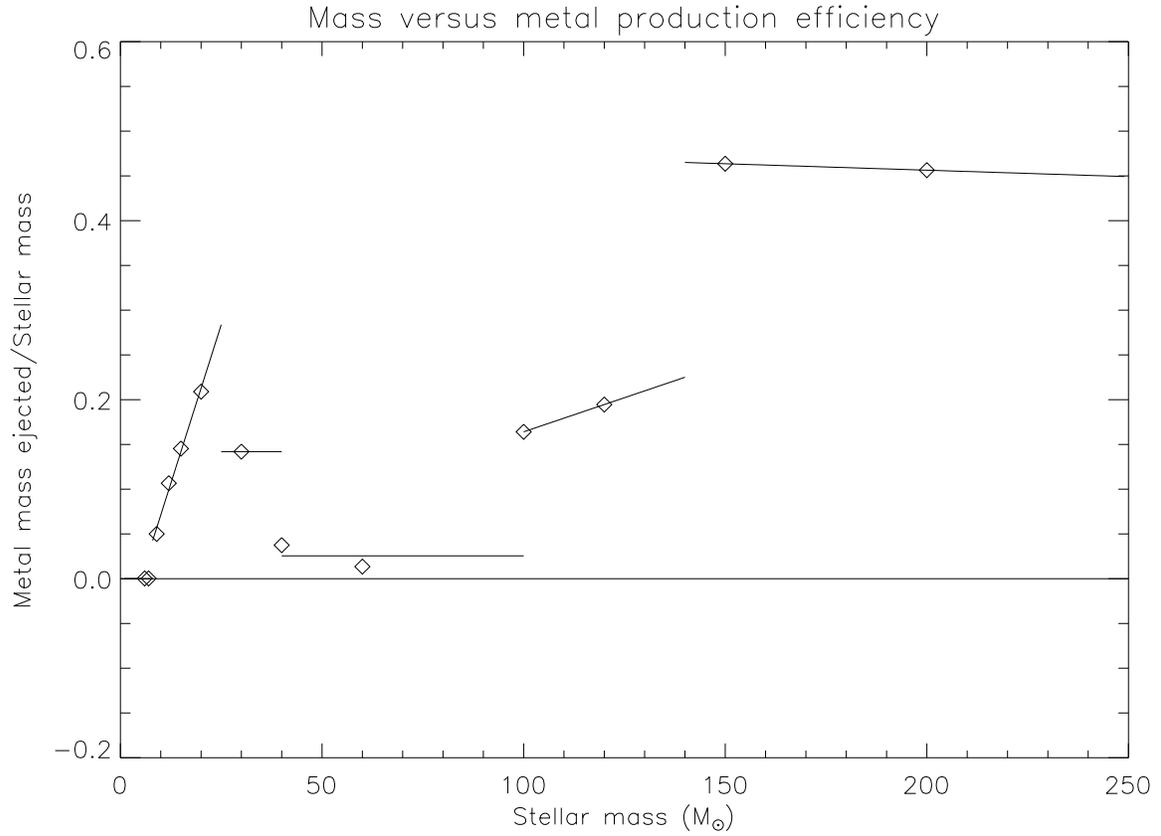}
\caption{%
The ratio of ejected stellar metals and stellar mass versus the mass of the star.  Note that each jump corresponds to a different stellar fate.  The lines correspond to the fitting functions given in Eq [\ref{eq:stellarmetals}].  Open diamonds correspond to the stellar models given in \citet{portinari/chiosi/bressan:1998}.   
}%
\label{fig:metals}
\end{figure}

The metal yields of stars with initial metallicity of $Z=1/50 Z_{\sun}$ were given in \citet{portinari/chiosi/bressan:1998}.  These models of metal production take into account stellar winds and supernova explosions.  The metal production efficiency (metal mass ejected from the star over initial stellar mass) is shown in Figure~\ref{fig:metals}.  It is clear that metal production depends strongly on the initial mass of the star and how the star ended its life.  
\begin{itemize}
\item From 6 to 8 $M_{\sun}$, the O/Ne/Mg core of the star collapses, or the star ejects its outer envelope, leaving a white dwarf or neutron star.
\item From 8 to 25 $M_{\sun}$, the iron core collapses, the star explodes as a supernova, and a neutron star is left as a remnant.  A significant amount of metals are ejected.
\item From 25 to 40 $M_{\sun}$, there is a weak supernova and a black hole is created by fallback.  The amount of metals that are ejected into the IGM decreases sharply, leaving most of the metals locked in the black hole.
\item From 40 to 100 $M_{\sun}$, the star directly collapses into a black hole.  The only metals produced are from mass loss during the star's life.
\item From 100 to 140 $M_{\sun}$, a pulsational pair instability supernova results.  This ejects the outer envelope of the star, and then the core collapses into a black hole.  Metals in the outer envelope pollute the IGM.
\item From 140 to 260 $M_{\sun}$, a pair instability supernova results, which completely disrupts the star and leaves no remnant.  All the metals are ejected into the IGM.
\item Above 260 $M_{\sun}$, the star collapses directly into a black hole, and there is no enrichment of the IGM.
\end{itemize}

Fitting functions were made for each of these mass intervals.  
\begin{equation}
  \label{eq:stellarmetals}
  \frac{M_{metals,ej}(m)}{m/M_{\sun}}=\left\{
    \begin{array}{ll}
 ~ 0, & < 8~{\rm M_{\sun}}\\
 -0.0708+0.0142  \frac{m}{M_{\sun}}, & 8-25 ~{\rm M_{\sun}} \\
 0.142 , & 25-40 ~{\rm M_{\sun}} \\
  0.0255, & 40-100 ~{\rm M_{\sun}} \\
  0.0117+0.00152\frac{m}{M_{\sun}}, & 100-140 ~{\rm M_{\sun}} \\
  0.485-0.000144\frac{m}{M_{\sun}}, & 140-260 ~{\rm M_{\sun}} 
 
  \end{array}       
  \right.
\end{equation}
$M_{metals,ej}$ is the mass in solar masses of ejected metals.  These fitting functions are limited because of the small number of data points being fit.  However, they should be able to serve to approximate the amount of metals formed in each mass range. For stars less massive than $8 M_\sun$, the amount of metals is negligible, and is assumed to be zero.  When pairing these metal production rates with an IMF, they agree well with \citet{giroux/shapiro:1996}.


Combining this with an initial mass function, we are in the position to be able to predict how many metals are formed by these first stars in a certain redshift range.  The amount of metals formed from stars is given by
\begin{equation}
\label{equ:rhosolve}
\rho_{metals}= \frac{\rho_*}{m_*}\int^{m_2}_{m_1}f(m)M_{metals, ej}(m)dm
\end{equation}
where $\rho_{metals}$ is the mass density of metals from a population of stars, $\rho_*$ is the stellar mass density, and $m_*$ is the mean stellar mass of the mass function (Eq.~[\ref{eq:mstar}]).  

The very first stars were metal free.  We will assume that metal free star formation only occurs to a certain critical metallicity, $Z=10^{-4} Z_{\sun}$.   \citep{schneider/etal:2002, bromm/larson:2004}. After this metallicity is reached, metal poor stars begin to form.  If this transition happened before a redshift of 15, all of the NIRB is from metal poor stars, and the maximum amount of metals would be produced.  Would these metal poor stars overproduce metals that we observe in the universe today?

We take the upper limit of metals that can be produced in the universe to be the solar metallicity, $Z=Z_{\sun}=0.02$.  Using $\Omega_b=0.044$ and $H_0=70\; \rm{kms^{-1}Mpc^{-1}}$, the baryonic density, $\rho_b$, is $6.0\times10^9\; \rm{M_\sun Mpc^{-3}}$.  Then the maximum metal mass density that can be produced is then
\begin{equation}
\rho_{metals}^{max}=\rho_b Z_{\sun}
\end{equation}
This gives a metal mass density of $\rho_{metals}^{max}=1.2\times10^8 \; \rm{M_{\sun}Mpc^{-3}}$.  If the predicted metallicity from Eq.~[\ref{equ:rhosolve}] exceeds this value, the model must be ruled out.

A population of metal poor stars that form from z=7-15 will form metals according to Eq.~[\ref{equ:rhosolve}].  This depends on the mass function and the star formation rate.  The amount of metals formed for the upper and lower $1-\sigma$ limit of $\dot{\rho}_*$ for each mass function are shown in table \ref{deltat}.  We have found that a population of metal poor stars do {\it not} overproduce metals that we observe in the universe today, except for the Larson mass function upper $1-\sigma$ value for the star formation rate.    

In a recent paper by \citet{dwek/arendt/krennrich:2005}, they showed if the entire NIRB excess is from stars, these stars would produce about 87 \% of the metal and helium abundance we see today.  This would leave little room for production in more recent eras from Pop II and Pop I stars.  In order to avoid this, they stated that stars can collapse directly into black holes, locking up their metals and therefore not ejecting any metals into the IGM.  They used this arguement to suggest that the NIRB must come from very massive Pop III stars, which are massive enough to collapse directly into black holes.  We showed that this is not necessarily the case: metal poor stars with a regular mass function can also produce less than the observed amount of metals.  By doing the calculations using the stellar data that takes into account both ejected metals and remnants, we find that the first stars may only contribute as little as less than 1 \% of the metals in the IGM, which is certainly feesable.



\begin{table}[t]
\begin{center}
\begin{tabular}{|l|l|l|l|}
\hline
IMF       &  $\dot{\rho}_*$ &$\rho_{metals} \; M_{\sun}Mpc^{-3}$ \\ 
\hline
Salpeter & 0.48 &   $ 9.59 \times 10^{5}$\\
 & 12 &$2.40 \times 10^7 $\\
Larson & 0.30 &$1.61 \times 10^7 $\\
 & 7.6 & $4.09 \times 10^8 $ \\

\hline
\end{tabular}
\caption{%
   The mass density of metals being produced by metal poor stars from a redshift of 7 to 15.  Values of $\dot{\rho}_*$ are the upper and lower 1-$\sigma$ limits given in Table \ref{tab:generation}.  All cases (except for the Larson mass function with an upper 1-$\sigma$ limit for $\dot{\rho}_*$) give values of $\rho_{metals}$ that is less that the  metal density we see today ($\rho_{metals}=1.2 \times 10^8 \; M_\sun Mpc^{-3}$), indicating that metal poor stars can form up to $z=7$ and not overproduce metals.  
}%
\label{deltat}
\end{center}
\end{table}

%
We now have a picture of how star formation could have occurred in the early universe.  Metal free stars began forming early.  As they die, they pollute the IGM with metals.  Once a critical metallicity is reached, metal poor stars begin to form.  These metal poor stars can continue to form up to a redshift of 7 without overproducing metals in the IGM, while still providing enough ultraviolet photons to make up the Near Infrared Background.

\section{AMPLITUDE OF MATTER FLUCTUATIONS AT SMALL SCALES}
\label{sec:remarks}


In the previous work on the near infrared background from
early stars \citep{santos/bromm/kamionkowski:2002,salvaterra/ferrara:2003,madau/silk:2005},
it was concluded that a substantial fraction of {\it collapsed} baryons in the universe
must have been converted into stars.
In \S~\ref{sec:sfr2} we have considered a fraction of {\it all} baryons in the universe that 
was locked up in stars at a given $z$, and now it is natural for us to ask how many
baryons collapsed in dark matter halos were converted into stars.
Of course, this fraction, $\eta$, which is sometimes called the ``star formation efficiency'', 
must not exceed unity. 

In Appendix~\ref{sec:rhodot}, we derive an analytical model for the cosmic star formation
rate (Eq.~[\ref{eq:sfranal}]) which has often been used in the literature:
\begin{equation}
  \dot{\rho}_*(z) 
  \simeq 0.536~M_\sun~{\rm Mpc}^{-3}~{\rm yr^{-1}}~
  \left(\frac{\eta}{0.1}\right)
\left(\frac{1+z}{10}\right)^{3/2}
  y_{\rm min}(z) e^{-y_{\rm min}^2(z)/2},
  \label{eq:sfranal2}
\end{equation}
where 
\begin{equation}
  y_{\rm min}(z) \simeq 
  \frac{1.68(1+z)}{\sigma[M_{\rm min}(z)]},
\end{equation}
and $\sigma(M)$ is the present-day r.m.s. amplitude of mass fluctuations given by
\begin{equation}
  \sigma^2(M) = \int_0^\infty 
  \frac{k^2dk}{2\pi^2}P(k)\left[\frac{3j_1(kR_M)}{kR_M}\right]^2.
\end{equation}
Here, $P(k)$ is the linear power spectrum of density fluctuations at present,
$j_1(x)=\sin(x)/x^2-\cos(x)/x$ is the spherical Bessel function of order 1,
$R_M$ is a radius defined by $M=(4\pi/3)\overline{\rho}_{m0}R_M^3$, and 
$M_{\rm min}(z)$ is the minimum mass of dark matter halos which can host
star formation. 
For illustration purposes, let us assume that star formation occurs when
cooling via atomic hydrogen becomes efficient, 
$M> M_{\rm min}(z)\simeq 10^8~M_\sun[(1+z)/10]^{-3/2}$ \citep{barkana/loeb:2001}.
This mass scale roughly corresponds to the wavenumber of 
$k=\pi/R_M\gtrsim 36.8[(1+z)/10]^{1/2}~{\rm Mpc}^{-1}$. 

By comparing this analytical model to the observational data, one can constrain
$\eta$ and/or $\sigma[M_{\rm min}(z)]$. The usual approach is to
constrain $\eta$ by fixing $\sigma[M_{\rm min}(z)]$ with $P(k)$ extrapolated from 
$k\lesssim 1~{\rm Mpc}^{-1}$ 
(which corresponds to $M\gtrsim 5\times 10^{12}~M_\sun$)
down to much smaller scales, $k\gtrsim 36.8[(1+z)/10]^{1/2}~{\rm Mpc}^{-1}$,
for which we do not have any direct observational constraints yet. 
This is potentially a dangerous approach. Eq.~[\ref{eq:sfranal2}]
implies that $\dot{\rho}_*(z)$ is exponentially sensitive to $\sigma[M_{\rm min}(z)]$,
and a slight increase in $\sigma[M_{\rm min}(z)]$ may substantially reduce $\eta$ for a given 
$\dot{\rho}_*(z)$, and thus one must not ignore the fact that we do not know
the precise value of $\sigma[M_{\rm min}(z)]$ without relying on extrapolations
of $P(k)$ by a factor of more than 40 in $k$.
Therefore, the stellar origin of the near infrared background cannot be ruled out
on the basis of $\eta$ until we understand the amplitude of matter fluctuations
on small scales.
One may reverse the argument: it might be possible to explore the small-scale
fluctuations by using the near infrared background, for a given star formation efficiency.



\section{CONCLUSIONS}
\label{sec:conclusion}

We have presented detailed theoretical calculations of 
the intensity of the cosmic near infrared background from early stars. 
We have shown that the intensity is essentially determined by the mass-weighted
mean nuclear burning energy of stars (for a given mass spectrum of early stars)
and the cosmic star formation rate. 
The prediction is not sensitive to stellar metallicity (Table~\ref{tab:nuInu}),
while uncertainty from the initial mass spectrum could be large, as we have very little
knowledge about the form of mass spectrum for early stars.  Our simple analytical calculations agree well with recent numerical calculations of the spectrum using the CLOUDY code \citep{dwek/arendt/krennrich:2005}.
The measured intensity at $1-2~\mu{\rm m}$ can be used to infer the cosmic star 
formation rate at $7<z<15$, which is difficult to constrain by other means.
Although the current data are quite uncertain due 
to subtraction of the zodiacal light, the inferred star formation rate,
$\dot{\rho}_*\sim 0.3-12~{\rm M_{\odot}~yr^{-1}~Mpc^{-3}}$ at $z=7-15$
for $m>5~M_\sun$,
is significantly higher than the low-$z$ rates.
We have shown that this does not exclude the stellar origin of the cosmic
near infrared background, as it merely requires more than 
0.016--0.49\% of baryons to be processed in stars at any given time between 
$z=7$ and 15 (depending on metallicity and initial mass spectrum; 
see Table~\ref{tab:generation}). 
Such a high star formation rate at high $z$ may be consistent with recent theoretical
proposals \citep{cen:2003,mackey/bromm/hernquist:2003}.  In addition, the derived star formation rate does not overproduce metals in the IGM (unless using the upper 1-$\sigma$ value paired with a Larson mass function), and may produce as little as less than 1 \% of the metals in the IGM.  
More accurate determination of the near infrared background is absolutely 
necessary to yield a meaningful estimate of the star formation rate with any 
confidence. 
If the future data demand too high of a star formation rate 
for the stellar origin to be viable, then
other sources that might contribute to the near infrared 
background, such as early quasars, should be invoked
\citep{cooray/yoshida:2004,madau/silk:2005}.

\acknowledgments
We thank K. Ahn, M. Alvarez, V. Bromm, N.J. Evans II, A. Ferrara, 
J. Scalo, P.R. Shapiro and N. Yoshida for 
valuable comments on early versions of the paper and fruitful discussions. 

\appendix
\section{DERIVATION OF VOLUME EMISSIVITY}
\label{sec:emissivity}

In this Appendix we derive the formula for volume emissivity,
$p(\nu,z)$, which is formally given by
\begin{equation}
  \label{eq:org}
  p(\nu,z)  = \sum_\alpha
     \int_{m_1}^{m_2} dm \frac{dn_*(m,z)}{dm} \overline{L}^\alpha_\nu(m),
\end{equation}
where $\alpha$ labels various radiative processes (e.g., line), 
$\overline{L}^\alpha_\nu$ is a time-averaged energy spectrum of relevant 
emission (luminosity in frequency interval [$\nu$, $\nu+d\nu$]), 
and $dn_*/dm$ is the comoving number density of stars {\it that are alive} 
at a given $z$ in mass interval [$m$, $m+dm$]. 
Subtlety exists as we need to properly take into account dead stars.
(Since dead stars do not contribute to the emissivity, they must be removed
from Eq.~[\ref{eq:org}].)
The goal of this Appendix is to derive the formula for $p(\nu,z)$ 
which is properly corrected for dead stars.

\subsection{Case with No Dead Stars}

First, as the simplest example (and for illustration purposes)
let us derive the formula which does not 
correct for dead stars. (Note that we do not use this formula. The correct
formula will be given in the next subsection.)
We write $dn_*/dm$ as
\begin{equation}
  \frac{dn_*(m,z)}{dm} 
  = n_*(z) f(m),
\end{equation}
where $f(m)$ is a probability distribution function of stellar masses
(also known as the mass spectrum) normalized to unity for a certain mass range
such that 
\begin{equation}
\int_{m_1}^{m_2} dm~f(m)=1.
\end{equation}
We assume that $f(m)$ is independent 
of time. For example, $f(m)\propto m^{-2.35}$, the Salpeter
mass spectrum, is independent of time. In principle, however,
$f(m)$ may depend on time when there is a characteristic
stellar mass scale (e.g., Larson's mass spectrum)
that increases or decreases with time. 
One may expect $m_c$ to decrease as the metal enrichment proceeds,
for example. Nevertheless, we shall assume that $m_c$ is independent
of time, at least for the redshift range that we consider.

We may use the comoving mass density of stars, $\rho_*(z)$, instead of the 
comoving number density, $n_*(z)$. The relation is
\begin{equation}
  \rho_*(z) = m_* n_*(z),
\end{equation}
where $m_*$ is the mean stellar mass given by
\begin{equation}
  m_* \equiv \int_{m_1}^{m_2} dm~mf(m).
  \label{eq:mstar}
\end{equation}
Using this, the volume emissivity (not corrected for dead stars)
becomes
\begin{equation}
  p(\nu,z)  = \frac{\rho_*(z)}{m_*}
     \sum_\alpha \int_{m_1}^{m_2} dm f(m) \overline{L}^\alpha_\nu(m).
  \label{eq:psf}
\end{equation}
\citet{salvaterra/ferrara:2003} used a version of this 
Eq.\footnote{\citet{salvaterra/ferrara:2003}
actually used $p(\nu,z)  = \rho_*(z)
\sum_\alpha \int_{m_1}^{m_2} dm f(m) \overline{L}^\alpha_\nu(m)/m$,
which is off by a factor of $m/m_*$.}, and therefore
they did not correct their emissivity for dead stars.

\subsection{Emissivity Corrected for Dead Stars}

Now, we correct emissivity for dead stars by simply removing 
them from $n_*(z)$:
\begin{equation}
  \frac{dn_*(m,z)}{dm} 
  = f(m) \int_{T_0(z)-\tau(m)}^{T_0(z)} dt~\dot{n}_*,
\end{equation}
where 
$\dot{n}_*$ is a rate of star formation,
$T_0(z)$ is the time between when the universe started 
forming stars and the time corresponding to $z$,
and $\tau(m)$ is a stellar main-sequence lifetime.
Eq.~[\ref{eq:org}] then becomes
\begin{equation}
  p(\nu,z)  = \sum_\alpha
  \int_{m_1}^{m_2}dm~f(m)\overline{L}_\nu^\alpha(m)
  \int_{T_0(z)-\tau(m)}^{T_0(z)}\frac{dt~\dot{\rho}_*}{m_*}.
  \label{eq:emformula}
\end{equation}
This result may be simplified when the stellar
lifetime is much shorter than $T_0(z)$ (which is about the same as the age
of the universe at $z$). As we have shown in Figure~1,
the main-sequence lifetime of stars contributing to the near infrared 
background is always shorter than the age of the universe.
We Taylor expand the integral over time to obtain the final formula:
\begin{equation}
  p(\nu,z)  
\simeq \frac{\dot{\rho}_*(z)}{m_*} \sum_\alpha
\int_{m_1}^{m_2}dm~f(m)\overline{L}_\nu^\alpha(m)\tau(m).
\label{eq:p}
\end{equation}

\section{STELLAR MASS DENSITY}
\label{sec:omegastar}

\subsection{Case with No Dead Stars}
\label{sec:density_nodead}

{\it Cumulative} mass density of stars that were formed 
at $t_1<t<t_2$ is given by
\begin{equation}
 \rho^{cum}_*(t_1,t_2) 
 = \int_{t_1}^{t_2} dt~\dot{\rho}_* 
 \approx (t_2-t_1)\dot{\rho}_*,
\label{eq:cum}
\end{equation}
where we have assumed that $\dot{\rho}_*$ is approximately constant over 
$t_1<t<t_2$. Using $z$ instead of $t$, one gets
\begin{eqnarray}
 \nonumber
 \rho^{cum}_*(z_1,z_2) 
 &=& \int_{z_1}^{z_2} \frac{dz~\dot{\rho}_*}{H(z)(1+z)} \\
 \nonumber
 &\approx& \frac1{\Omega_m^{1/2}H_0}\int_{z_1}^{z_2}  \frac{dz~\dot{\rho}_*}{(1+z)^{5/2}} \\
 \nonumber
 &\approx& 5.509\times 10^8~{\rm M_\sun~Mpc^{-3}}
 \left(\frac{0.14}{\Omega_mh^2}\right)^{1/2}
 \left[\frac{\dot{\rho}_*}{\rm M_\sun~yr^{-1}~Mpc^{-3}}\right]\\
 & &\times \left[\left(\frac{10}{1+z_1}\right)^{3/2}-\left(\frac{10}{1+z_2}\right)^{3/2}\right],
\end{eqnarray}
where we have assumed that $z_1\gg (\Omega_\Lambda/\Omega_m)^{1/3}-1\simeq 0.3$
so that dark energy contribution to $H(z)$ can be ignored.
This is cumulative density, as it includes those stars which had already died.
The fractional stellar mass density relative to the critical density,
$\Omega_*$, is then given by
\begin{eqnarray}
\nonumber
  \Omega^{cum}_*(z_1,z_2) 
  &\equiv& \frac{\rho_*(z_1,z_2)}{\rho_{c0}}\\
  \Omega^{cum}_*(z_1,z_2)h^2 &\approx& 1.985\times 10^{-3}
  \left(\frac{0.14}{\Omega_mh^2}\right)^{1/2}
 \left[\frac{\dot{\rho}_*}{\rm M_\sun~yr^{-1}~Mpc^{-3}}\right]\\
 & &\times \left[\left(\frac{10}{1+z_1}\right)^{3/2}-
\left(\frac{10}{1+z_2}\right)^{3/2}\right],
 \end{eqnarray}
where $\rho_{c0}=2.775h^2\times 10^{11}~{\rm M_\sun~Mpc^{-3}}$ 
is the present-day critical density. 

\subsection{Stellar Mass Density Corrected for Dead Stars}
\label{sec:density_dead}

We remove dead stars (whose lifetime, $\tau(m)$, is much shorter
than the Hubble time) from the stellar mass density to obtain
\begin{equation}
  \rho_*(t_1,t_2) \approx  \dot{\rho}_*
  \int_{m_1}^{m_2}dm~f(m)\tau(m).
\end{equation}
Comparing this with the cumulative stellar density (Eq.~[\ref{eq:cum}]), 
one finds the relation
\begin{equation}
  \rho_*(t_1,t_2) \approx \frac{\rho^{cum}_*(t_1,t_2)}{N_{gen}},
\end{equation}
where $N_{gen}$ is the average number of generation of stars:
\begin{equation}
  N_{gen} = \frac{t_2-t_1}{\int_{m_1}^{m_2}dm~f(m)\tau(m)}.
\end{equation}

\section{ANALYTICAL MODEL OF THE COSMIC STAR FORMATION RATE}
\label{sec:rhodot}

A popular assumption usually made for an analytical model of the cosmic star
formation rate at high $z$ is that $\dot{\rho}_*(z)$ is related to the mass function
of dark matter halos:
\begin{equation}
  \dot{\rho}_*(z) 
  = \eta\overline{\rho}_{b0}H(z)(1+z)
  \frac{\partial F[M>M_{\rm min}(z)]}{\partial z},
\end{equation}
where $\overline{\rho}_{b0}=2.775\times 10^{11}~(\Omega_bh^2)~M_\sun~{\rm Mpc}^{-3}$
is the present-day mean baryon density, and $\eta$ represents a 
``star formation efficiency'', a constant fraction of baryonic gas in dark matter halos 
that was converted into stars. It is admittedly too simplistic to assume that 
$\eta$ is independent of halo mass or redshift. 
For example, negative feedback from a star forming in a single mini-halo
might prevent the formation of multiple stars in the same halo, which
would imply $\eta\propto M^{-1}$. Therefore, this parameterization
of the efficiency serves merely as an order-of-magnitude representation of 
the true efficiency.
Finally, $F$ is a ``collapse fraction'', 
a fraction of mass in the universe collapsed into halos more massive
than $M_{\rm min}(z)$,
\begin{equation}
  F[M>M_{\rm min}(z)]
 = \frac1{\overline{\rho}_{m0}}\int_{M_{\rm min}(z)}^\infty dM \frac{dn_h(M,z)}{dM}M,
\end{equation}
where $\overline{\rho}_{m0}=2.775\times 
10^{11}~(\Omega_mh^2)~M_\sun~{\rm Mpc}^{-3}$
is the present-day mean total mass density, and $M$ 
is the halo mass (not to be confused with the stellar mass, $m$),
and $dn_h/dM$ is the halo mass function.
The Press-Schechter mass function gives the collapse fraction in terms
of the complimentary error function,
\begin{equation}
  F[M>M_{\rm min}(z)]
  = 
  \sqrt{\frac{2}{\pi}}\int_{y_{\rm min}(z)}^\infty dy~e^{-y^2/2},
\end{equation}
where
\begin{equation}
  y_{\rm min}(z) \equiv \frac{\delta_c}{\sigma[M_{\rm min}(z)]D(z)}
  \approx \frac{1.68(1+z)}{\sigma[M_{\rm min}(z)]},
\end{equation}
in the redshift range of interest ($z>7$).
One thus finds
\begin{eqnarray}
 \nonumber
  \frac{\partial F[M>M_{\rm min}(z)]}{\partial z}
  &=&
 \sqrt{\frac{2}{\pi}}
 \frac{dy_{\rm min}}{dz}
 e^{-y_{\rm min}^2(z)/2}\\
 \nonumber
  &\approx&
  \sqrt{\frac{2}{\pi}}
  \frac{y_{\rm min}(z)}{1+z}
    \left[
 1 + \frac{d\ln M_{\rm min}(z)}{d\ln (1+z)}
 \left.\frac{d\ln \sigma^{-1}}{d\ln M}\right|_{M=M_{\rm min}(z)}
 \right]e^{-y_{\rm min}^2(z)/2}\\
  &=&
  \sqrt{\frac{2}{\pi}}
  \frac{y_{\rm min}(z)}{1+z}
 \frac{1- n_{\rm eff}[M_{\rm min}(z)]}4   e^{-y_{\rm min}^2(z)/2}.
\end{eqnarray}
Here, we have assumed that $M_{\rm min}(z)$ is proportional
to the virial mass and thus $M_{\rm min}(z)\propto (1+z)^{-3/2}$,
and defined an effective slope of the power spectrum as 
$[n_{\rm eff}(M)+3]/6 \equiv d\ln\sigma^{-1}/d\ln M$.
At the scale of interest, the effective slope lies between
$-2.5$ and $-2.8$; thus, the prefactor, $(1- n_{\rm eff})/4$,
is of order unity and the exact value is not so important.
Setting it to be unity, we obtain a fully analytical formula for the star 
formation rate at high redshifts,
\begin{eqnarray}
  \nonumber
  \dot{\rho}_*(z) 
  &\approx& \sqrt{\frac{2}{\pi}}\eta\overline{\rho}_{b0}H(z)
  y_{\rm min}(z) e^{-y_{\rm min}^2(z)/2}\\
  \nonumber
  &=& 
  0.536~M_\sun~{\rm Mpc}^{-3}~{\rm yr^{-1}}~
  \left(\frac{\eta}{0.1}\frac{\Omega_bh^2}{0.02}\right)
\left(\frac{\Omega_mh^2}{0.14}\right)^{1/2}\\
& &\times 
\left(\frac{1+z}{10}\right)^{3/2}
  y_{\rm min}(z) e^{-y_{\rm min}^2(z)/2}.
\label{eq:sfranal}
\end{eqnarray}


\end{document}